\newif\ifacmartsty
\acmartstyfalse
\newif\ifcameraready
\camerareadyfalse
\documentclass[sigconf,10pt]{sig-alternate-10pt}
\ifacmartsty
\else
\fi

\usepackage{amsmath,amsfonts}
\usepackage[T1]{fontenc}
\usepackage[utf8]{inputenc}
\usepackage[numbers]{natbib}

\usepackage{upquote}

\usepackage{microtype}
\UseMicrotypeSet[protrusion]{basicmath} % disable protrusion for tt fonts

\usepackage{graphicx,grffile}
\usepackage{times} 
\usepackage{algorithm, algpseudocode}
\usepackage{multirow}
\usepackage{multicol}
\usepackage{xspace}
\usepackage{tabularx}
\usepackage{ragged2e}
\usepackage{booktabs}
\usepackage{paralist}
\usepackage[american]{babel}
\usepackage[shortlabels]{enumitem}
\usepackage{courier,color,wrapfig}
\usepackage{balance}
\usepackage{enumitem}
\usepackage{epstopdf}
\usepackage{booktabs}
\usepackage{url}
\usepackage{pifont}
\usepackage{subfigure}
\usepackage{tikz}
\usepackage{mathtools}
\usepackage[normalem]{ulem}

\usepackage{etoolbox}
\usepackage{ifthen}
\usepackage[textsize=tiny]{todonotes}
\usepackage{tikz}
\usetikzlibrary[calc,intersections,through,backgrounds,arrows,positioning,decorations.pathmorphing,fit]
\usepackage[compact]{titlesec}
\usepackage{caption}
\makeatletter
\def\maxwidth{\ifdim\Gin@nat@width>\linewidth\linewidth\else\Gin@nat@width\fi}
\def\maxheight{\ifdim\Gin@nat@height>\textheight\textheight\else\Gin@nat@height\fi}
\makeatother
\setkeys{Gin}{width=\maxwidth,height=\maxheight,keepaspectratio}
\makeatletter
\g@addto@macro{\UrlBreaks}{\UrlOrds}
\makeatother

\makeatletter
\let\origsection\section
\let\origsubsection\subsection
\let\origsubsubsection\subsubsection

\renewcommand\section{\@ifstar{\starsection}{\nostarsection}}
\renewcommand\subsection{\@ifstar{\starsubsection}{\nostarsubsection}}
\renewcommand\subsubsection{\@ifstar{\starsubsubsection}{\nostarsubsubsection}}

\newcommand\sectionprelude{\vspace{-0.5ex}}
\newcommand\sectionpostlude{\vspace{-1ex}}

\newcommand\subsectionprelude{\vspace{-0.5ex}}
\newcommand\subsectionpostlude{\vspace{-0.5ex}}

\newcommand\subsubsectionprelude{\vspace{-0ex}}
\newcommand\subsubsectionpostlude{\vspace{-0.5ex}}

\newcommand\nostarsection[1]{\sectionprelude\origsection{#1}\sectionpostlude}
\newcommand\starsection[1]{\sectionprelude\origsection*{#1}\sectionpostlude}

\newcommand\nostarsubsection[1]{\subsectionprelude\origsubsection{#1}\subsectionpostlude}
\newcommand\starsubsection[1]{\subsectionprelude\origsubsection*{#1}\subsectionpostlude}

\newcommand\nostarsubsubsection[1]{\subsubsectionprelude\origsubsubsection{#1}\subsubsectionpostlude}
\newcommand\starsubsubsection[1]{\subsubsectionprelude\origsubsubsection*{#1}\subsubsectionpostlude}

\makeatother
\newcommand\paraspace{\vspace*{0.5ex}}
\providecommand\parab[1]{\noindent{\textbf{#1}}}
\providecommand\parae[1]{\noindent{\textit{#1}}}

\ifacmartsty
\fi

\apptocmd\normalsize{%
\abovedisplayskip=5pt
\abovedisplayshortskip=5pt
\belowdisplayskip=5pt
\belowdisplayshortskip=5pt
}{}{}

\ifacmartsty
\renewcommand\footnotetextcopyrightpermission[1]{} % removes footnote with conference info
\setcopyright{none}
\acmConference[ACM MobiSys'22]{Submitted to MobiSys'22}{TBA}{2022} 
\fi

\algnewcommand\INPUT{\item[\textbf{Input:}]}%
\algnewcommand\OUTPUT{\item[\textbf{Output:}]}%

\savebox\strutbox{$\vphantom{\dfrac10}$}

\newcommand{\Sysname}{\textsc{AUTOCAST}\xspace}
\newcommand{\sysname}{\textsc{autocast}\xspace}

\DeclareMathDelimiter{(}{\mathopen} {operators}{"28}{largesymbols}{"00}
\DeclareMathDelimiter{)}{\mathclose}{operators}{"29}{largesymbols}{"01}

\newcommand{\etc}{\textit{etc.}\xspace}
\newcommand{\ie}{\textit{i.e.,}\xspace}
\newcommand{\eg}{\textit{e.g.,}\xspace}

\newcommand{\secref}[1]{\S\ref{#1}}
\newcommand{\figref}[1]{Figure~\ref{#1}}

\newcommand{\eqnref}[1]{Equation~\ref{#1}}

\begin{document}

% \title{\sysname: Autonomous Vehicle Multicast Scheduling for Scalable Perception Augmentation}
% \title{\sysname: Scalable and Efficient 3D Sensor Sharing between Autonomous Vehicles}
\title{AutoCast: Scalable Infrastructure-less Cooperative Perception for Distributed Collaborative Driving}

% Scalable Cooperative Perception
% Coordination for Scalable Cooperative Perception

\numberofauthors{6}

\author{
\alignauthor
Hang Qiu
\email{hangqiu@usc.edu}
\alignauthor
Po-Han Huang
\email{pohanh@usc.edu}
\alignauthor
Namo Asavisanu
\email{namo@usc.edu}
\and
\alignauthor
Xiaochen Liu
\email{liu851@usc.edu}
\alignauthor
Konstantinos Psounis
\email{kpsounis@usc.edu}
\alignauthor
Ramesh Govindan
\email{ramesh@usc.edu}
}
% \affiliation{University of Southern California}
% \author{
% {Hang Qiu}
% \\
% University of Southern California
% \and
% {Po-Han Huang}
% \\
% University of Southern California
% \and
% {Xiaochen Liu}
% \\
% University of Southern California
% \and
% {Konstantinos Psounis}
% \\
% University of Southern California
% \and
% {Ramesh Govindan}
% \\
% University of Southern California
% } % end author

% \author{ }

% \author{Hang Qiu}
% \affiliation{%
% 	\institution{University of Southern California}
% }
% \email{hangqiu@usc.edu}

% \author{Po-Han Huang}
% \affiliation{%
% 	\institution{University of Southern California}
% }
% \email{pohanh@usc.edu}

% \author{Xiaochen Liu}
% \affiliation{%
% 	\institution{University of Southern California}
% }
% \email{liu851@usc.edu}

% \author{Konstantinos Psounis}
% \affiliation{%
% 	\institution{University of Southern California}
% }
% \email{kpsounis@usc.edu}

% \author{Ramesh Govindan}
% \affiliation{%
% 	\institution{University of Southern California}
% }
% \email{ramesh@usc.edu}

% \renewcommand{\shortauthors}{H. Qiu, P. Huang, X. Liu, K. Psounis and R. Govindan}

%\keywords{Autonomous driving, augmented reality, V2V communications, multicasting scheduling, Markov decision process}
\maketitle
\begin{abstract}
  Autonomous vehicles use 3D sensors for perception. Cooperative perception enables vehicles to share sensor readings with each other to improve safety. Prior work in cooperative perception  scales poorly even with infrastructure support. \sysname enables scalable infrastructure-less cooperative perception using direct vehicle-to-vehicle communication. It carefully determines which objects to share based on positional relationships between traffic participants, and the time evolution of their trajectories. It coordinates vehicles and optimally schedules transmissions in a distributed fashion.  Extensive evaluation results under different scenarios show that, unlike competing approaches, \sysname can avoid crashes and near-misses which occur frequently without cooperative perception, its performance scales gracefully in dense traffic scenarios providing 2-4x visibility into safety critical objects compared to existing cooperative perception schemes, its transmission schedules can be completed on the real radio testbed, and its scheduling algorithm is near-optimal with negligible computation overhead. 
\end{abstract}

\section{Introduction} 
\label{s:introduction}

Autonomous driving technology has made great strides in transforming our daily transportation. To be socially acceptable and widely deployed, a next challenge is to ensure dependability over a broad set of unusual traffic situations and
corner cases~\cite{teslaaccident}, exceeding human driving safety levels (100M miles between fatalities~\cite{FARS}). To achieve this requires successfully navigating unusual events and corner cases (\eg objects on the roadway, a pedestrian crossing highway \etc)
that one might encounter in billions of miles of driving.

% \parab{Autonomous driving: a brief primer.}
% Today's autonomous vehicles attempt to achieve these goals using sensors that provide \textit{depth perception}. An example of such a sensor is the LiDAR, which works by emitting laser pulses and estimating depth of objects in the scene from the time to receive reflected laser beams. For example, a 64-beam Velodyne LIDAR~\cite{velodyne64} has a rotating assembly that contains 64 laser beams. The LiDAR completes 10 rotations per second. Each rotation generates a LiDAR frame, which contains a collection of points (reflected beams) forming a \textit{point cloud}. Each point is represented by its three-dimensional position, and optionally other attributes (such as intensity). A \textit{perception} software component in an autonomous vehicle uses the sensor data to accurately position the vehicle with respect to the surrounding environment, and to recognize road hazards as well as the road driving surface. A \textit{planner} component uses this information to generate \textit{trajectories} (a sequence of \textit{waypoints}, together with times at which the vehicle is expected to be at each waypoint). The vehicle's \textit{controller} actuates the mechanical and electrical components of the vehicle to effect these trajectories.

% \parab{Line-of-sight limitations}.
The reliability of today's autonomous driving solutions is critically dependent on the accuracy of its perception component. However, 3-D sensors like LiDAR and stereo cameras suffer from line of sight limitations: other vehicles and traffic participants (pedestrians, bicyclists) can block a vehicle's view. For example, across the nuScenes dataset~\cite{nuscenes}, 53.14\% of over 204K labeled pedestrians and 59.43\% of 483K annotated vehicles are occluded. The blind spots caused by occlusion can compromise the reliability of object detection and downstream path planning in many driving situations, including left turns, lane changes and overtaking.

\begin{figure}[t]
\centering
\vspace{-1mm}
\includegraphics[width=\columnwidth]{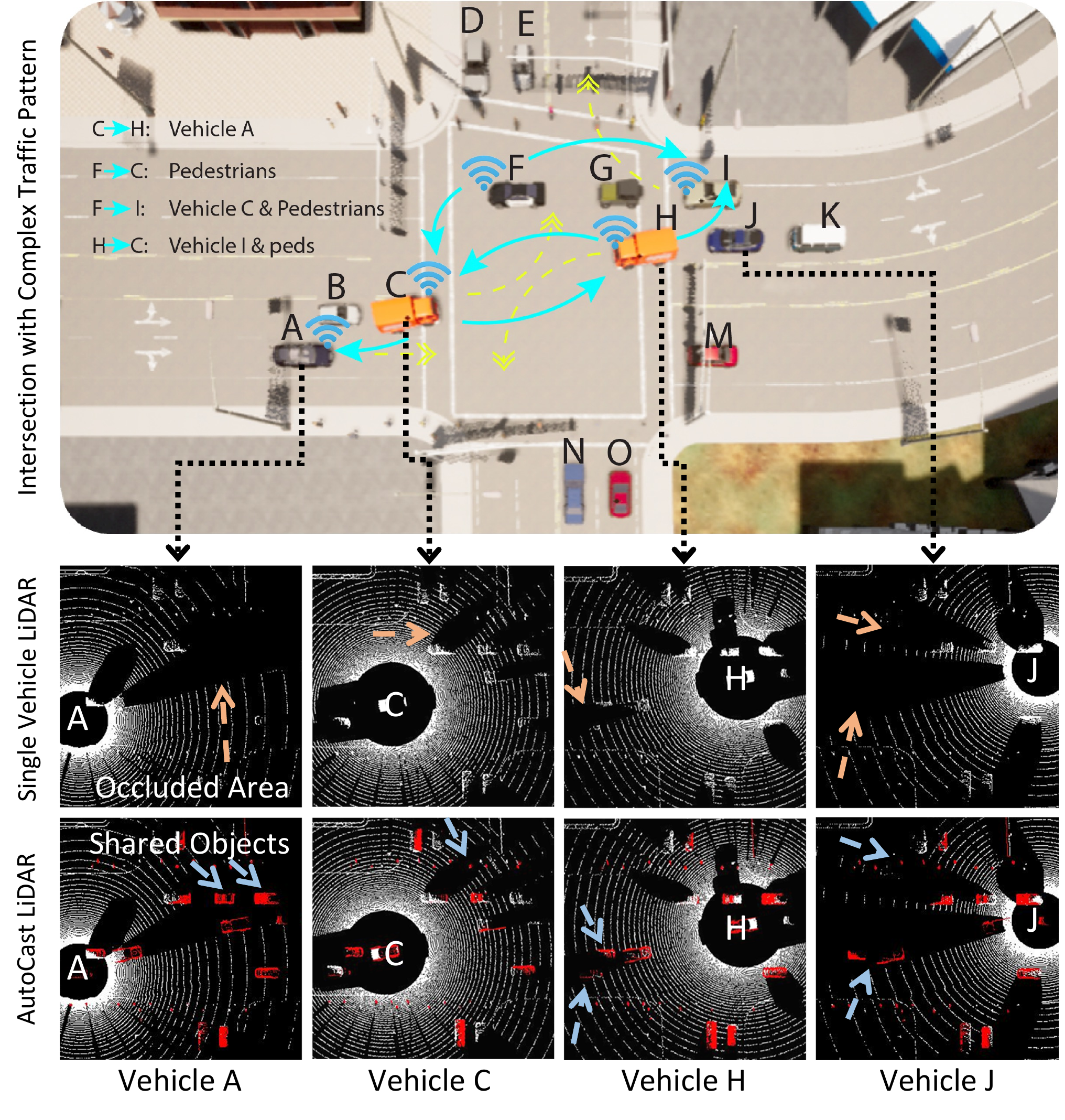}
\vspace{-8mm}
\caption{\sysname enables multi-vehicle cooperative perception in a busy intersection. The top graph shows that \sysname orchestrates vehicles to selectively share useful information (light blue arrows) about occluded objects, which the receiver vehicles cannot see, but may affect the receivers trajectories (yellow dashed arrow). The bottom graph shows the LiDAR perspectives of vehicle A, C, H, J: the upper row shows the invisible area (orange arrows) before sharing; the lower row shows the previous invisible objects (red points, blue arrows) visible after sharing.}
\label{fig:motiv}
\vspace{-2mm}
\end{figure}
\parab{Cooperative perception.}
To address this limitation, recent work~\cite{AVR,v2vnet, chen2019fcooper,Gomes:VNC, emp} has proposed a novel new direction, \textit{cooperative perception}. In this approach, vehicles cooperatively exchange sensor readings from 3D sensors to extend their visual horizon (\figref{fig:motiv}). The benefits of cooperative perception for autonomous driving systems are clear: a vehicle can make decisions much earlier than it otherwise might have been able to. Cooperative perception is a natural next step beyond earlier prior work that considered safety enhancements by having vehicles actively broadcast their location continuously over short range DSRC radios~\cite{DSRC}; these approaches cannot capture passive participants (pedestrians and bicyclists), as cooperative perception can.

\parab{Connected vehicles: enabling cooperative perception.}
Prior work leverages recent advances in vehicular communication to enable cooperative perception.
AVR~\cite{Qiu:Mobisys} enables vehicles to \textit{directly} exchange stereo camera point clouds via
\textit{vehicle-to-vehicle (V2V)} communication.
EMP~\cite{emp} exploits infrastructure support to share non-overlapping segments of LiDAR point clouds via \textit{vehicle-to-infrastructure (V2I)} communication, using an edge server as a \textit{relay}.
Using static infrastructure, V2I can achieve higher bandwidth compared to direct mobile V2V communication. For example, V2I using LTE~\cite{LTE} can achieve nominal bandwidths of up to 300~Mbps~\cite{LTESpeed,LTESpeed1,LTESpeed2}, whereas 
% V2V via DSRC\footnote{Dedicated short range communication, a V2V  standard.}~\cite{80211p} can achieve 3-27~Mbps. 
current commercial V2V products using DSRC\footnote{Dedicated short range communication, a V2V  standard.}~\cite{80211p} tend to achieve up to 6~Mbps.
LTE and Wi-Fi, which many modern vehicles have, increasingly support \emph{direct} modes and wider channel bandwidths. However, WiFi-direct via 802.11n/ac or 60 GHz (ad) cannot adapt to the highly variable wireless channel between fast moving vehicles.
% , so the only 802.11 version used today for vehicular communications is still 802.11p (DSRC). 
% LTE-direct can support 10-20 simultaneous transmissions at 20~Mbps between vehicles within a 1 \(km^2\) area per 10 MHz channel, if communicating vehicles are within 100m~\cite{LTEDirectOverview}. 
Existing  LTE-direct~\cite{LTE-Direct} chips for vehicular applications, tend to achieve around 10~Mbps\footnote{5GAA~\cite{5gaa} extensions for vehicles potentially  reach higher rates.}. 

\parab{Sharing point clouds is desirable but challenging.} Similar to EMP and AVR, we advocate for sharing raw points for cooperative perception, as opposed to  processed information, such as bounding boxes or visual features. Sharing bounding boxes caps the cooperative perception by the accuracy of the object detector deployed on the transmitter vehicle. Among popular object detectors (VoxelNet~\cite{voxelnet}, PointPillars~\cite{pointpillars}, CenterPoint~\cite{centerpoint}), the mean average precision (mAP) can vary by 8\%.
The other alternative is to use a neural network to extract and share features. This approach would \textit{constrain innovation}: the receiver might wish to use the shared feature for many different purposes (object detection, drivable space segmentation, trajectory planning), and the transmitted features might limit the efficacy of these tasks. 
Even for the purpose of object detection, prior work
has shown that early fusion of shared point clouds can result in \textit{higher accuracy} compared to late fusion of processed features~\cite{v2vnet}. 
% This is because, increasingly, deep neural networks process visual information, and they can learn more from raw data than from processed representations or hand-crafted features. 

Nevertheless, sharing raw points is challenging. While AVR and EMP has demonstrated the feasibility, their scale is limited by the network bottleneck: AVR is limited to two vehicles using V2V communication; EMP scales up to six vehicles via V2I, but needs infrastructure support.

\parab{Our focus: scalable and  infrastructure-less sharing.}
In this paper, we explore the problem of \textit{scalable infrastructure-less cooperative perception}, which permits vehicles to share raw sensor data in dense traffic scenarios without the dependency on edge servers. To motivate the problem, consider a busy intersection (\figref{fig:motiv}) with complex traffic dynamics where no infrastructure support is available: people and bicyclists crossing the street, traffic waiting to make a turn, together with traffic flowing in the direction of the green light. In such a scenario, there may be tens to hundreds of traffic participants; if each vehicle could still share information about participants that it sees, other vehicles would have more complete information to plan better trajectories. 
Without an edge server, the challenges lie in two folds: fitting the shared data into the narrower V2V bandwidth; coordinating and scheduling transmissions to avoid packet collision.

% \parab{The network is the bottleneck.}
% There is a significant mismatch between the rates at which 3D sensors generate data, and what vehicular communication standards can sustain. A 64-beam LiDAR can generate 10 \textit{point clouds}/sec totaling 2.2 M points/sec \cite{velodyne64}, which generates data at 2.2~Gbps. However, as discussed above, V2V standards like DSRC and LTE-direct can likely support a few Mbps to a few tens of Mbps. This would seem to suggest that exchanging raw sensor information between vehicles is infeasible, at least for the foreseeable future.

\parab{Sharing point clouds via V2V channel is feasible.}
The closest prior work, EMP~\cite{emp}, demonstrated that it is possible to transfer, through V2I channel, upto six non-overlapping segments of point clouds (each of size 30-38 KB). 
For the purpose of autonomous driving and collision avoidance, instead of an entire segment, a vehicle can transmit point clouds of relevant objects in the scene.
By emulating a 64-beam Velodyne LiDAR (which generates 2.2 M points/sec~\cite{velodyne64}) in photorealistic simulations of different driving scenarios (see \secref{sec:eval}), we have found that point clouds for a detected object (\eg a truck close by) have up to 200 points (38.4~kbits\footnote{LiDAR beams are radial, points are denser for nearby objects than for faraway ones. Objects faraway have smaller point cloud as well.}, an order of magnitude smaller than EMP segments).
Ideally, to enable a vehicle to track participants (especially fast-moving vehicles) precisely, each vehicle must receive (and make trajectory planning decisions on) point clouds at sub-second timescales. The finest \textit{decision interval}, denoted by $T_d$ is 100~ms, which is the interval at which the Velodyne LiDAR generates a frame \cite{velodyne64}. 
% Every $T_d$, then, each vehicle transmits point clouds to other vehicles. 
At 10 Hz (as clocked in EMP, when $T_d=100 ms$), a nominal V2V channel capacity of 10~Mbps can fit from 25 dense vehicle point clouds to hundreds of sparser (faraway) or smaller objects (pedestrians and cyclists). Compression can further increase that number.
%

% Thus, scalable cooperative perception using LiDAR point cloud is at the \textit{edge of feasibility}: the number of participants at a busy intersection can potentially be larger than that, so cooperative perception must use other techniques (discussed below) to scale to these settings.

\parab{Ad hoc coordination and scheduling.} Enabling cooperative perception with infrastructure support can be a huge investment with limited coverage. For example,  EMP~\cite{emp} supports cooperative perception for six vehicles every 100 m, which requires a dedicated edge service every 100 m. Deploying such services in urban scenarios, either to regional road-side units (RSUs) or to aggregated edge clusters (\eg cellular towers), can be very expensive in terms of compute and cost. Moreover, using edge servers as relays unnecessarily duplicates transmissions, which wastes already scarce wireless bandwidth, and increases end-to-end (V2V) delivery latency.
In areas not covered by such edge services,  there is no scalable way to enable cooperative perception. 
Instead, we argue for an ad hoc approach.
With the discovery range of LTE-direct~\cite{LTE_whitepaper} ranging from 170 m (urban) to 550 m (rural), it is feasible for vehicles to coordinate and form clusters on demand, which can significantly reduce the deployment cost.

% In addition to using V2V channels, another challenge that the absence of 

\parab{Contributions.} In this paper, we discuss the design and implementation of \sysname, a system that enables scalable infrastructure-less cooperative perception. 
% Because cooperative perception using LiDAR point clouds is on the edge of feasibility, \sysname uses several techniques to maximize the utility of information transmitted on the wireless channel. 
% Like AVR, it only transmits point clouds of dynamic objects.
Beyond extracting object points, \sysname uses several techniques to maximize the utility of information transmitted on the wireless channel. It determines \textit{visibility} and \textit{relevance} when deciding whether to transmit. For example, if car \textbf{A} wants to transmit roadway objects to car \textbf{B}, some of those objects may already be visible to \textbf{B}; \textbf{A} need only transmit roadway objects \textit{occluded} at \textbf{B}. Moreover, \textbf{B} may not need all occluded roadway objects, since some of them may not be \textit{relevant} to its driving decision. 
% For example, if it is currently driving straight, it may not need occluded objects on the opposite lanes on a divided highway. 
\sysname uses these criteria to determine which objects should be transmitted when.

\begin{figure}
\centering
%\vspace{-1mm}
\includegraphics[width=0.8\columnwidth]{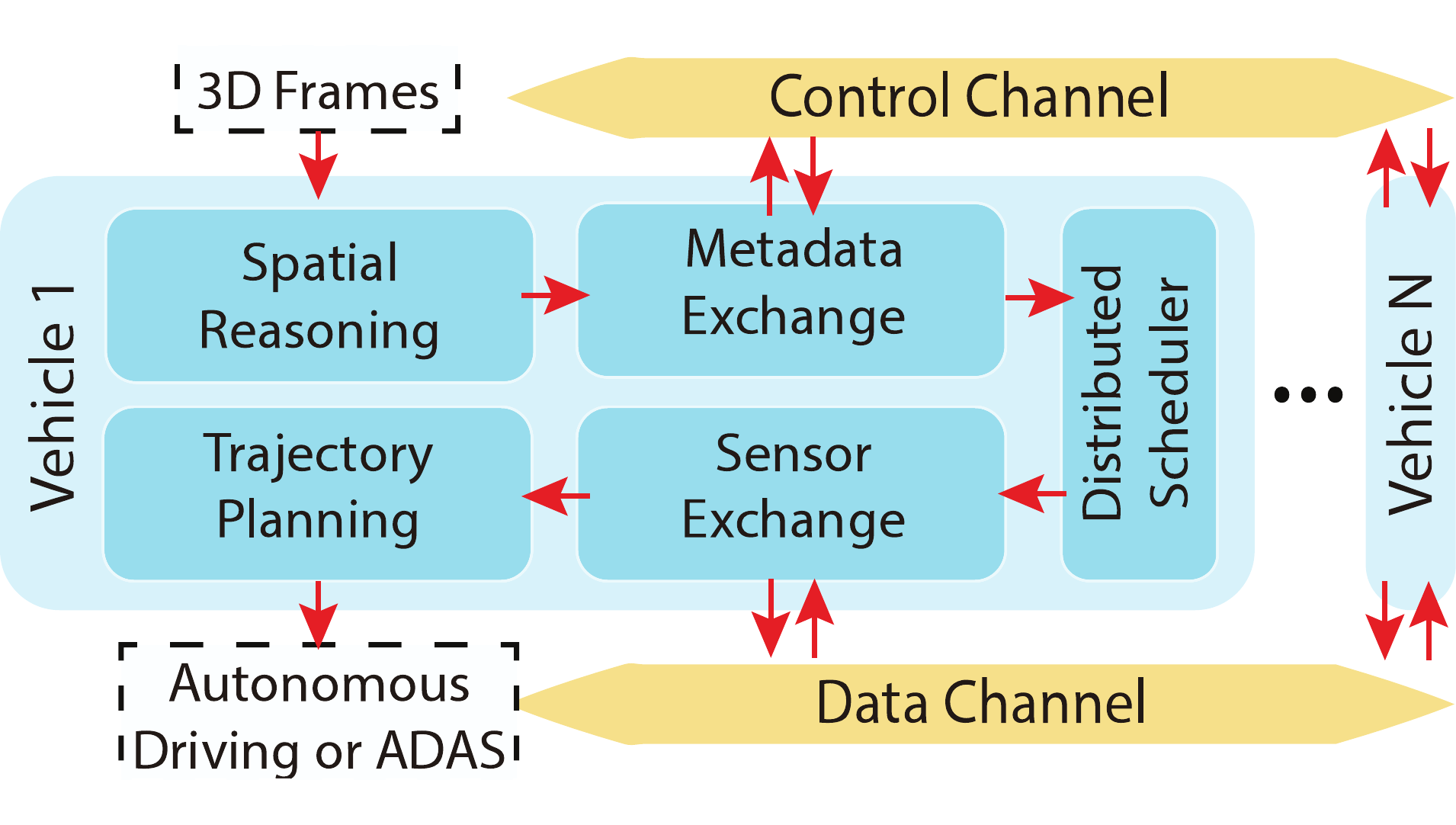}
\vspace{-3mm}
\caption{\sysname system architecture}
\label{fig:sysarch}
\vspace{-1mm}
\end{figure}

\paraspace \noindent To this end, \sysname makes three contributions.

\vspace{-2mm}
\begin{itemize}[leftmargin=4mm]
\item A suite of fast \textit{spatial reasoning} algorithms that analyzes point clouds to determine visibility and relevance. 
\vspace{-2mm}
% At a high-level these rely on metadata exchanges between vehicles which share positions and trajectories of observed objects, and each vehicle uses fast geometric algorithms to analyze these object relationships.

\item
An efficient scheduling algorithm and a near-optimal heuristic that prioritizes safety critical transmissions. 
\vspace{-2mm}
% The MDP can reason about optimal ordering of objects across different radio frames, which is useful because \sysname schedules sensor data every 100~ms (the rate at which LiDAR frames are generated) which can in general encompass more than one radio frame. Because the optimal MDP solution requires dynamic programming which can be too slow in our setting, we also develop a greedy heuristic that we show to yield a near optimal solution for the experiments we consider.
\item
A planner that fuses shared points, estimates object motion, and finds collision free trajectories in a receding horizon.
% Third, to demonstrate the benefits of \sysname's cooperative perception end-to-end, we needed a trajectory planner for vehicles (so we could demonstrate crash-avoidance, for example). However, existing planners are all designed based on sensor data collected from a single vehicle. We developed a planner capable of reasoning cooperative perception; this planner extends $A\star$ search to find collision free trajectories.

\end{itemize}
\vspace{-2mm}

% \hangq{revisit after revising MDP section, add distributed coordination part potentially}

% \hangq{revisit after adding EMP results}
\noindent We implement \sysname in Carla \cite{CarLA}, a photo-realistic autonomous driving simulator.
Our evaluation results (\secref{sec:eval}) show that \sysname can reduce all 100\% hazardous  situations (crashes, deadlocks) caused by occlusion in single-vehicle based perception, significantly reduce near-miss cases by providing early situational awareness and increasing reaction time. The scheduling architecture prioritizes most relevant information, scales gracefully upto 40 vehicles within sharing range, raising visibility into safety critical objects by 2-4x of the time, 2-8x in visible size, avoiding all collisions that are inevitable using alternative baselines. We have also implemented DSRC-radio based prototype for coordinated transmissions. Experiments validate the DSRC's channel capacity to meet schedule on time. Finally, we have optimized the end-to-end \sysname pipeline to operate at >10~fps.

%%% Local Variables:
%%% mode: latex
%%% TeX-master: "main"
%%% End: 

\section{\Sysname Architecture}
\label{sec:overview}

\sysname's end-to-end architecture (\figref{fig:sysarch}) contains a \textit{control-plane} (\secref{s:control-plane}) that exchanges beacons and makes transmission scheduling decisions, and a \textit{data-plane} (\secref{s:data-plane}) that processes, transmits, and uses point clouds to make trajectory planning decisions. 
This decoupling of data and control ensures that bandwidth intensive point cloud data is directly transmitted between vehicles with \textit{minimum delay for real time decisions}, while at the same time the control plane is able to make near optimal scheduling decisions.

\parab{The control plane.} 
Two subcomponents constitute the control plane. The \textit{metadata exchange} component (\secref{sec:protocol}) 
% runs both on the vehicle and the  road-side-unit (RSU)
% \footnote{Without RSUs, \sysname can be extended to a distributed fashion with a leader-selection protocol among vehicles. In this paper, we assume an RSU available, leaving the distributed version for future work.}, and 
implements a protocol to exchange metadata (needed for scheduling, obtained from the data plane) among vehicles. The \textit{scheduler} (\secref{sec:scheduler}) uses this information to compute a transmission schedule,
% , relayed back to the vehicles using the metadata exchanger. 
which is  executed by the data plane.

\parab{The data plane.}
Two subcomponents constitute the data plane.  \textit{Spatial reasoning} (\secref{sec:assess}) extracts moving objects from LiDAR sensors. For each object, it determines which vehicles cannot see this object and to which subset of them the object would be relevant; those are the vehicles to whom this object should potentially be sent. Each vehicle runs a \textit{trajectory planning} component (\secref{sec:planning}), which fuses the receives objects to adapts its current trajectory.

\begin{figure}
\centering
\includegraphics[width=0.5\columnwidth]{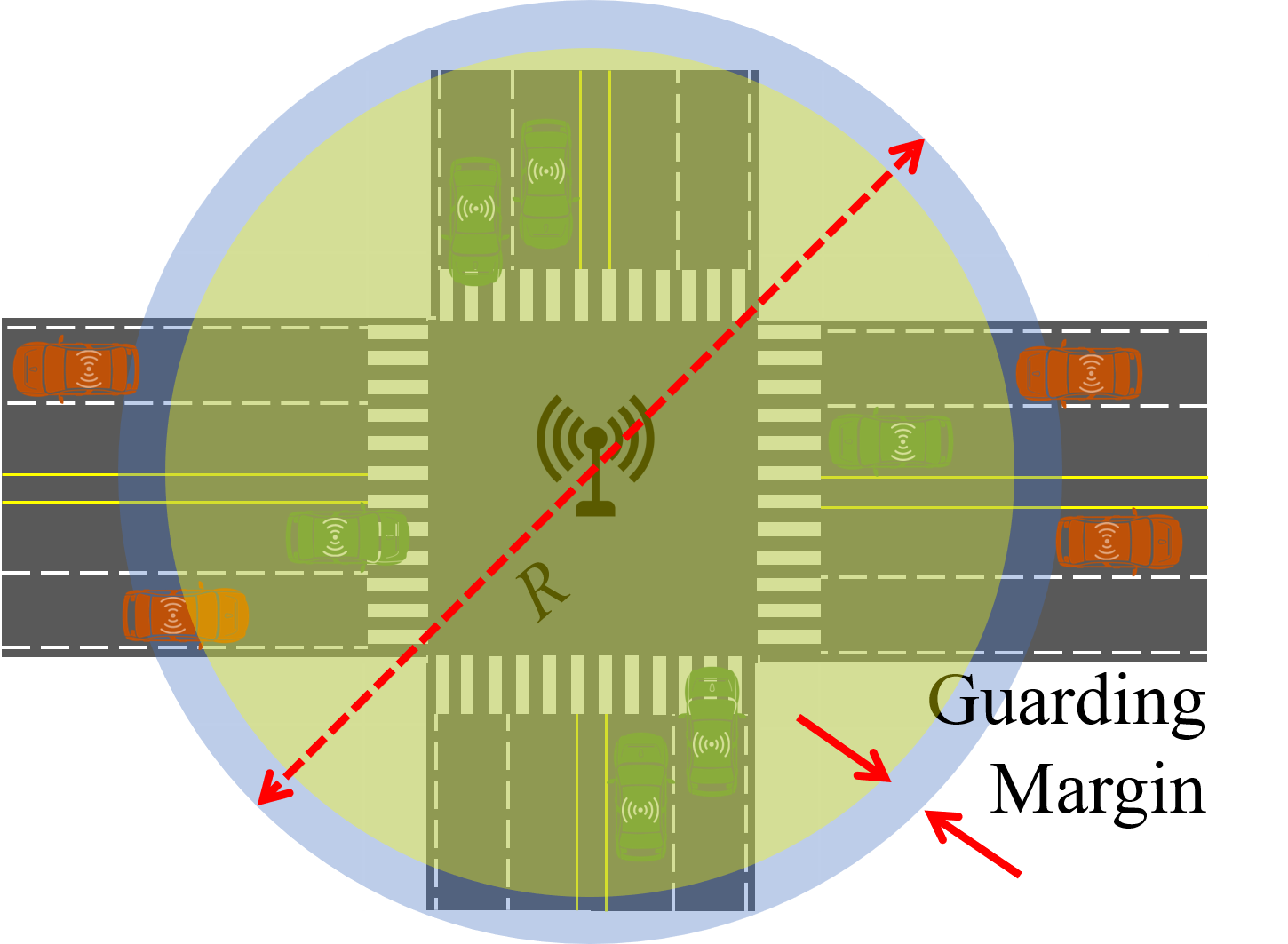}
\vspace{-2mm}
\caption{\textit{Scheduler domain} (idealized): green vehicles are selected participants; red vehicles are excluded.}
\label{fig:protocol_scenario}
\centering
% \vspace{-0.5cm}
\end{figure}

\section{Control Plane}
\label{s:control-plane}

We now describe metadata exchange and scheduling components (\figref{fig:sysarch}). Both the control plane and the data plane assume that each vehicle is able to accurately position itself using a 3-D map~\cite{CarMap} and Simultaneous Localization and Mapping (SLAM~\cite{vloam,loam}), so that each vehicle knows its own position precisely at all times. This kind of positioning technology is mature enough and has been widely deployed. 
% that all autonomous vehicles on the road today have it.

\subsection{Metadata Exchange}
\label{sec:protocol}

% \kostas{I recommend you create a figure which shows say an intersection, the cetner of the intersection, a circle of radius 50m around the intersectin, an RSU, a cluster head (if there is an RSU no need but the picture shoudl show both cases), the 100m distance frm two vechicles on the ends f the circle left and right, some guard gray area of say 2m inside the 50m circle 

% then you also create a plot on a highway, with 3   regions with channesl 1-2-1, where a guy becomes the leader who may not be in the center of a channel area (this is not needed) and then arund this leader you show a 50m circle again and cars within this circle.  this way reader realizes that the sharing regions are not the same with the channel regions. The channel regions are prespecified, wherease the exact boundaries of sharing regions depend on the location of a car the moment it becomes a leader.\figref{fig:protocol_scenario}}

%\kostas{after ramesh makes his pass we need to re-think the titles of this section and subsections and structure. the autocast system consists of three main pilars as discussed with ramesh, and the current data plane vs control plane structure doesn't do justice to what we have achieved}.

\parab{Deployment setting.}
Vehicles exchange metadata between themselves.
% and also with a road-side unit (RSU). 
Because the specifics of the metadata exchange can depend upon the relationship between radio range and road geometry, we describe metadata exchange for a concrete deployment setting, an \textit{intersection}. Intersections are also where cooperative perception can help most  \cite{Grembek:ArXiv}, because they are among the most hazardous parts of the road network.

% \begin{figure}
% \centering
% \includegraphics[width=0.7\columnwidth]{fig/Protocol}
% \caption{\sysname Protocol (Intersection w/ RSU).}
% \label{fig:protocol}
% \centering
% \vspace{-0.7cm}
% \end{figure}

\parab{Control messages.}
% Assume that there is a radio-equipped RSU at an intersection (\figref{fig:protocol_scenario}). 
\sysname participants 
% (the RSU and vehicles) 
periodically broadcast \textit{control messages} every $T_d$, the timescale at which the scheduler makes decisions (\secref{s:introduction}). These control messages are highly likely to reach each vehicle that is close to or at the intersection (\figref{fig:protocol_scenario}).
% , as well as the RSU. 
This is because lane widths are on the order of 3-4~m~\cite{USTest, EuropeTest}, so intersections of major streets with 3 lanes in each direction can be on the order of 30~m$\times$30~m. On the other hand, the nominal radio range of LTE-direct, denoted as $R$, is over 170~m (urban non-line-of-sight)~\cite{LTE_whitepaper, Qualcomm:LTE-Direct}, so even vehicles far away from the intersection can hear these messages. 
These control messages, or beacons, inform vehicles of their neighbors location, so that the scheduler can  identify all those vehicles within a circle of radius\footnote{\sysname associates a guard band  to allow for vehicle movement. This guard band can be calculated from the posted speed limits: at 40~mph, a vehicle can move $\sim1.8~m$ in one $T_d$ (100~ms) interval.} ($R/2$); these are the vehicles that can plausibly hear each other and they constitute the scheduler's \textit{domain} (\figref{fig:protocol_scenario})\footnote{In practice, the radio range may be irregular. A smaller domain radius (\eg 100 m, much less than maximum radio range $R$) can be chosen to ensure all participants can plausibly hear each other.}. 
% For these vehicles, using the received object maps, the RSU can compute a transmission schedule that determines which vehicle $i$ should transmit which object $k$ to which vehicle $j$. The RSU then broadcasts this transmission schedule to all vehicles. The precise form of the transmission schedule and the mechanism for transmitting this depends on the underlying radio technology (\eg DSRC, LTE-V); 
We discuss this, together with how vehicles multiplex their transmissions to execute the schedule, in \secref{sec:scheduler}.

\parab{Information exchanged in control messages.} 
In \sysname, participants exchange two types of information in these messages. Standardization efforts have defined V2V messaging formats that exchange similar \textit{cooperative awareness messages}~\cite{CAM-ETSI}; we have left it to future work to design a standard-compliant message exchange.

\parae{Trajectory.}
Each vehicle transmits its current \textit{trajectory} to other participants; the vehicle's planner (\secref{sec:planning}) generates and updates the trajectory every $T_d$. A trajectory (denoted by $t_i$ for vehicle $i$) consists of a series of \textit{waypoints} and their associated \textit{timestamps}. Each waypoint indicates the position a vehicle expects to be at the corresponding timestamp. To limit control overhead, \sysname down-samples trajectories into connected line segments for sharing. The first waypoint in the trajectory is the current \textit{vehicle pose}. 
% Denote by $t_i$ the trajectory of the $i$-th vehicle.

\parae{The object map.}
Using its 3D sensor, each vehicle can extract point clouds of \textit{roadway objects}; these are stationary or moving objects (vehicles, pedestrians, cyclists) on the road surface. 
%Each point in the point cloud has an associated position. 
Denote by $o_{i,k}$ the $k$-th object in vehicle $i$'s view.
Now, vehicle $i$ receives broadcasted trajectories from other vehicles. Using spatial reasoning techniques described in \secref{sec:assess}, vehicle $i$ computes the following two quantities for each $o_{i,k}$ in its view:
%\begin{itemize}
%\item 
(1) $v_{(i,k),j}$ is a boolean value that indicates whether $o_{i,k}$ is visible to vehicle $j$.
%\item 
(2) $r_{(i,k),j}$ is a value that indicates whether $o_{i,k}$ is \textit{relevant} to $j$'s current trajectory $t_j$. We make the notion of relevance precise in \secref{sec:assess}.
%\end{itemize}
%
Vehicle $i$ then broadcasts an \textit{object map} that contains: (a) an ID for each object $o_{i,k}$, (b) the size of $o_{i,k}$ in bytes, (c) $v_{(i,k),j}$, and (d) $r_{(i,k),j}$. In \secref{sec:scheduler}, we explain how the scheduler uses these values to compute a transmission schedule.

% \parae{The transmission schedule.}
% The RSU at the intersection receives trajectories and object maps from vehicles within radio range. It identifies all those vehicles within a circle of radius\footnote{\sysname associates a small guard band (\figref{fig:protocol_scenario}) to allow for vehicle movement. This guard band can be calculated from the posted speed limits: at 40~mph, a vehicle can move $\sim1.8~m$ in one $T_d$ (100~ms) interval.} ($R/2$, where $R$ is the radio range); these are the vehicles that can plausibly hear each other and they constitute the scheduler's \textit{domain} (\figref{fig:protocol_scenario})\footnote{In practice, the radio range may be irregular. A smaller domain radius (much less than maximum radio range $R$) can be chosen to ensure all participants can plausibly hear each other.}. For these vehicles, using the received object maps, the RSU can compute a transmission schedule that determines which vehicle $i$ should transmit which object $k$ to which vehicle $j$. The RSU then broadcasts this transmission schedule to all vehicles. The precise form of the transmission schedule and the mechanism for transmitting this depends on the underlying radio technology (\eg DSRC, LTE-V); we discuss this, together with how vehicles within a sharing region multiplex their transmissions to cooperatively execute the centralized schedule, in \secref{sec:scheduler}.

\parab{Loss compensation.}
% We have described \sysname in an intersection setting. In \secref{sec:discussion} we discuss how to deploy a distributed \sysname for any setting, \eg in a highway setting.
Control messages can be lost;  In case of a loss, we reuse the trajectories from the most recent control message (evicting waypoints up to the current timestamp). 
% To update the object map, one can extrapolate object locations using motion forecasts (of detected objects from modern perception modules \cite{MarcoPavoneForecast,FAF,IntentNet}). Instead of sharing forecasts, 
\sysname also \textit{extrapolates} objects' current locations based on latest locations and timestamps. With extrapolated location, \sysname recalculates the visibility ($v_{(i,k),j}$) and relevance ($r_{(i,k),j}$) metrics to update the object map.

\parab{Scheduler domain and clusterization.} \figref{fig:protocol_scenario} shows one example scheduler domain at an intersection. Beyond that, scheduler domains can be predefined in high-definition maps. Using SLAM, vehicles can easily figure out which domain they belong to. To avoid inter-domain interference, neighboring domains use different channels. Because the road network naturally isolates the map into blocks, we only need to assign scheduler domains with alternating channels along the length of the road, while each domain covers the full width of the road. Autonomous vehicles radio boxes are equipped with multiple V2X antennas~\cite{ismartways}. When they are at the border, they can participate in both clusters at the same time and handover from one to the other following their moving direction. There is a large body of literature~\cite{cluster_survey,mobdhop,mpbc, energybasedclustering, maxmind} enabling distributed clustering in the context of mobile ad hoc networks. In this work, we focus on the novel aspect of the system design, and leave the application of the best fitting clustering protocols to future work.

\subsection{\Sysname Scheduler} \label{sec:scheduler}

In this section, we discuss \sysname's scheduler. Depending on the network bandwidth and the number and size of objects $o_{i,k}$ relevant to other vehicles, it may not be feasible to transmit all relevant objects before the next decision interval $T_d$. The scheduler decides which objects to transmit at every decision interval and in what order. 
For example, if an object is likely to cause an imminent collision, it must have a higher priority in the transmission schedule.
The underlying PHY layer may be able to transmit multiple PHY frames\footnote{The term ``frame'' used in this section refer to the time steps in a decision interval $T_d$, which is different from the lidar frame in \secref{s:data-plane}} during one decision interval; the scheduler must decide which objects to transmit in which frames. 
Each vehicle computes the schedule using the common list of control messages in its domain (\figref{fig:protocol_scenario}); each vehicle then broadcasts the specific object in the assigned PHY frame.
%\kostas{Sharing session duration dictated by application. lidar gets new image every 100ms. LTE frame duration depends on standard version. Std LTE 10ms, LTE-V proposed to be 100ms. In geenral we may have multiple LTE frames per sharing sessoin thus MDP good to have. If only one frame no need. Last, LTE schedule does allow us to say which objects to go to which frame if multiple frames available.}

\parab{Preliminaries: notation and PHY layer.}
Let $\mathcal{C} = \{1,...,C\}$ be the set of vehicles and $\mathcal{K} = \{1,...,K\}$ be the set of objects across all vehicles.  
% Let $n$ be the $n^{th}$ frame of the current decision interval which has a total of $N$ frames. 
Let $x^n_{i,k}=\{0, 1\}$ be a decision variable indicating whether vehicle $i$  transmits its object $k$ at frame $n$ (of N frames in total),
%$t^n_{i,v} \in \mathcal{R}^+$ be the transmission time for vehicle $i$ to transmit object $v$ at frame $n$, 
$S^n_f$ be the size of frame $n$,
and $T^n$ be the duration of frame $n$, thus
$\sum_n T^n \leq T^d$, and $T^n=S^n_f/B$ where $B$ is the bandwidth.

$T^n$ and $B$ depend on the PHY technology. There are two technologies available today, DSRC and LTE-V (a variant of LTE-direct).
$B$ varies between 5 and 10 Mbps and $T^n$ varies between 10 and 100 ms in current standards.
DSRC is based on TDMA while LTE-V offers both an OFDM option and a TDMA option. In case of OFDM, a frame multiplexes transmissions from multiple vehicles similarly to uplink frames in cellular networks. In case of TDMA, a frame consists of concatenated (in time) transmissions from multiple vehicles.
% We discuss more about these technologies in \secref{sec:discussion}. 

%\kostas{About the discussion in the discussion section: DSRC assumes TDMA across different transmitters via carrier sense. LTE-V has two options: TDMA, or OFDMA (that is, multiplexing in both time and frequency, where each vehicle transmits at a specified non colliding time-frequency blocks similar to cell phones in the uplink). In our paper the implementation of the scheduler has assumed the simpler case where TDMA is used across vehicles where vehicles are ordered in decreasing sum reward of their scheduled objects. Where should we state this? In the scheduling section, in the evaluation section, in the discussion section?}
%\ramesh{If the scheduler design can accommodate both TDMA and OFDMA, then we should (a) say that here, (b) discuss our use of the TDMA in the eval, and (c) discuss what complexity we would encounter in implementing OFDMA based transmissions in the discussion section. Does that make sense?}
%The rest of the discussion applies to both technologies and the whole range of speeds and frame durations.

%\parab{PHY layer.}
We assume the PHY layer uses QPSK as per common practice in vehicle communication systems due to the challenging channel \cite{3GPP:36.785,3GPP:36.885}.
Thus, the system can deliver $L = B \times \log(1+\gamma_{QPSK})$ bits per unit time, where $\gamma_{QPSK}$ is the SINR value required by QPSK.
We model the PHY layer by the  probability of successful delivery of an object between two vehicles. 
We define a $C\times C$ channel matrix comprising of these probabilities as follows, $\mathbb{P} = [p_{i,j}, i,j=1\ldots C]$,
%\begin{equation}
%\mathbb{P} = \begin{bmatrix} p_{11} & p_{12} & p_{13} & \dots  & p_{1C} \\ p_{21} & p_{22} & p_{23} & \dots  & p_{2C} \\ \vdots & \vdots & \vdots & \ddots & \vdots \\p_{C1} & p_{C2} & p_{C3} & \dots  & p_{CC} \end{bmatrix},  \label{eq:ChannelMatrix}
%\end{equation}
where $p_{i,j} \in [0,1]$ indicates the probability of delivery from vehicle $i$ to $j$, $\forall i, j \in \mathcal{C}$, ($p_{i,j}$ are assumed independent).
% In case of a packet loss, \sysname will use the most recent packet message available and extrapolate, see \secref{sec:eval} for more details.

%\todo{add a bit detail of trajectory extrapolation}
\parab{Problem formulation: markov decision process.} 
Because a decision interval may have multiple frames, we formulate the scheduling problem as a Markov Decision Process (MDP) such that it optimizes the scheduling  across all frames.

\parae{State.}
Let $h^n_{(i,k),j}$ indicate whether vehicle $j$ has received object $k$ from vehicle $i$ by frame $n$, and $q_j^n = \{h^n_{(i,k),j}, \forall i,k \}$.
We define the state of the system at frame $n$ by $S^n  = \{q^n_1, q^n_2, ..., q^n_C\}
% , \forall n \in \{1,...,N\}
$, where $S^n \in \mathcal{S}$, with $\mathcal{S}$ denoting the state space. Since the MDP state changes for each frame, $n$ represents the discrete time steps over which the MDP operates.

\parae{Action.}
Let $s_{i,k}$ denote the size of object $o_{i,k}$,
% \begin{align}
$ A^n = \{x^n_{i,k}=\{0, 1\}, \forall i,k
%\in \mathcal{C},\mathcal{K} 
| \sum_{\forall i \in \mathcal{C}}\sum_{\forall k \in \mathcal{K}}{s_{i,k}\times x^n_{i,k}} \leq S^n_f \}$ 
% \nonumber
% \end{align}
denote the action taken at time step $n$, where $A^n \in \mathcal{A}$, 
% with $\mathcal{A}$ denoting 
the action space. 
%Note that $s_{i,k}$ denotes the size of object $o_{i,k}$ and the constraint ensures that the selected objects fit in the frame.
%\kostas{Following up on my previous note, a reader may wonder here how can different vehicles/transmitters use the same frame? In traditional LTE as discussed above this is well known thanks to OFDMA. But when TDMA is used like in DSRC and the simpler version of LTE-V and in our implementation, one may argue that what we call a frame here is merely putting together in time multiple smaller frames each used by individual vehicles to transmit their own scheduled objects. Do you think this may create confusion to some readers and needs to be clarified?}
%\ramesh{Perhaps as a footnote so the reviewer understands that we are careful? See my reply above also.}

\parae{Reward function.}
To maximize the total rewards the system needs to carefully decide the action ($A^n$) based on the current state ($S^n$).
When the action is decided, the reward follows:
% \vspace{-1mm}
\small{
\begin{align}
R^n = \sum_{\forall j \in \mathcal{C}}\sum_{\forall i \in \mathcal{C}}\sum_{\forall k \in \mathcal{K}} x^n_{i,k} \times (h^{n+1}_{(i,k),j}-h^{n}_{(i,k),j}) \times y^n_{(i,k),j}~, \label{eq:reward_func}
\end{align}}
\normalsize
where $y^n_{(i,k),j}$ is the reward when object $k$ is transmitted from vehicle $i$ to vehicle $j$, which we define by
\[y^n_{(i,k),j} = (1 - v_{(i,k),j}) \times r_{(i,k),j}~.\] The rationale for this definition is that there is a reward if vehicle $j$ receives object $k$ by vehicle $i$ if object $k$ is invisible and relevant to vehicle $j$, see \secref{sec:protocol}.
%\kostas{reward can be binary (either relevant or non-relevant) or a number between 0 and 1 (e.g. we normalize with the time to collision to give higher relevance to more urgent objects). Also, if we integrate the object size/transmission time to the reward, then the normalized reward becomes non binary anyway. The boolean logic above implies that we have taken so far the stance of a binary reward...}

\parae{Transition probability.}
We compute the transmission probability from one state to another based on action ${A^n}$ as follows:
%\begin{align}
%&P_{S^n, S^{n+1}}^{A^n} = \nonumber\\
%&\prod_{\forall i: (i, k) \in \left\{(i, k) %\in \mathcal{C}\times\mathcal{K}|x^n_{i,k} = 1 %\right\} }\left(\prod_{\forall j \in %\mathcal{V}^1_{(i,k)}} p_{i,j} \times %\prod_{\forall j \in \mathcal{V}^0_{(i,k)}} %(1-p_{i,j})\right)~,
%\end{align}
% \vspace{-1mm}
% \small
\begin{align}
\scriptstyle
P_{S^n, S^{n+1}}^{A^n} = 
\prod_{\forall i: x^n_{i,k}=1 }\left(\prod_{\forall j \in \mathcal{V}^1} p_{i,j} \times \prod_{\forall j \in \mathcal{V}^0} (1-p_{i,j})\right)~,
\end{align}
\normalsize
where 
$\mathcal{V}^1 = \{ j |(h^{n+1}_{(i,k),j} = 1, h^{n}_{(i,k),j} = 0, x^n_{i,k} = 1)\} $ corresponds to all vehicles $j$ which successfully received the scheduled object during time step $n+1$,  
whereas $\mathcal{V}^0 = \{ j \in \mathcal{C}|(h^{n+1}_{(i,k),j} = 0, h^{n}_{(i,k),j} = 0, x^n_{i,k} = 1\} $ corresponds to vehicles which lost the scheduled object during that frame. 
%Hence, the value of $h^{n+1}_{(i,k),j}$ and $h^n_{(i,k),j}$ is used to represent all the possible combinations of events with respect to which subset of vehicles successfully receive a transmitted object. 
%Note that the condition $(h^{n+1}_{(i,k),j} = 1$, $h^n_{(i,k),j} = 1)$ is not of interest to us because this will not increase the reward, and the occurrence of $(h^{n+1}_{(i,k),j} = 0$, $h^n_{(i,k),j} = 1)$ is impossible.
%\kostas{the above combinatorial set is not carefully expressed, talk to po han and fix}

\parae{Markov Decision Process.}
We define a finite-horizon MDP by the tuple $\mathcal{M}(\mathcal{S},\mathcal{A},P_{S^n, S^{n+1}}^{A^n},R^n)$.
To solve the MDP, we first define a policy $\pi$
to be a mapping from states to actions and seek to find the optimal policy which maximizes the (expected, discounted) sum of the rewards occurring from the selected actions over a (potentially infinite) time horizon. 

To find the optimal policy a recursive approach is used, which updates (i) policy decisions $\pi(S^n)$ at each frame $n$ and (ii) the value function $\mathcal{U}^\pi(S^n)$ which keeps track of the sum of the rewards if policy $\pi$ is followed from state $S^n$. Specifically, the corresponding recursive formula is given by:
% \vspace{-1mm}
% \small
\begin{align} \label{eq:Bellman}
\scriptstyle
\mathcal{U}^{\pi(S^n)}(S^n) = 
% \sum_{\forall S^{n+1} \in \mathcal{S}}P_{S^n, S^{n+1}}^{\pi(S^n)} \cdot 
\sum_{\forall S^{n+1}}P_{S^n, S^{n+1}}^{\pi(S^n)} \cdot 
\left(R^n
% (S^n, S^{n+1}, \pi(S^n))
+\mathcal{U}^{{\pi}(S^{n+1})}(S^{n+1}) \right).
\end{align}
% \small
% \begin{align} \label{eq:Bellman}
% &\mathcal{U}^{\pi(S^n)}(S^n) = 
% % \nonumber\\
% &\sum_{\forall S^{n+1} \in \mathcal{S}}P_{S^n, S^{n+1}}^{\pi(S^n)} \cdot 
% \left(R^n
% % (S^n, S^{n+1}, \pi(S^n))
% +\mathcal{U}^{{\pi}(S^{n+1})}(S^{n+1}) \right).
% \end{align}
\normalsize
The rationale behind this equation is that an optimal policy can be constructed by going backwards in time: we first construct an optimal policy for the tail subproblem corresponding to the last step at time step $n=N$, then apply the optimal policy to the tail subproblem corresponding to the last two steps at time step $n=N-1$, and continue in this manner until an optimal policy for the entire problem is formed.

%\begin{align}
%&\mathcal{U}^{{\pi}^{*}(S^n)}(S^n) = \nonumber\\
%&\max_{\forall A^n \in \mathcal{A}}\left\{\sum_{\forall S^{n+1} \in \mathcal{S}}P_{S^n, S^{n+1}}^{{\pi}^{*}(S^n)} \cdot \left(R^n(S^{n}, S^{n+1}, A^n)+\mathcal{U}^{{\pi}^{*}(S^{n+1})}(S^{n+1}) \right)\right\}.
%\end{align}
One may use dynamic programming (DP) to find the optimal policy $\pi^*(S^n) \in \mathcal{A}$ 
which maximizes the value function in \eqnref{eq:Bellman}
using the Bellman equation~\cite{MDP}.
However, the time complexity grows exponentially~\cite{Powell:ADP} with the number of states. This motivates us to seek more scalable approaches.

\parab{Scheduling algorithms.} 
We start by defining the weight of an object at frame $n$, denoted by $H^n_{i,k}$, to be the total rewards gained by the system if
the object is successfully delivered to all interested vehicles:
\vspace{-1mm}
% \small
\begin{equation} 
\scriptstyle
H^n_{(i,k)} = \sum_{\forall j \in \mathcal{C}, i \neq j}{y^n_{(i,k),j}} \times (1 - h^{n-1}_{(i,k),j})~. \label{eq:GreedyMetric}
\end{equation}
\normalsize
Note that the term $(1 - h^{n-1}_{(i,k),j})$ indicates the object has not been received during previous frames.

\begin{figure}[!bht]
\centering
\vspace{-4mm}
\includegraphics[width=0.5\columnwidth]{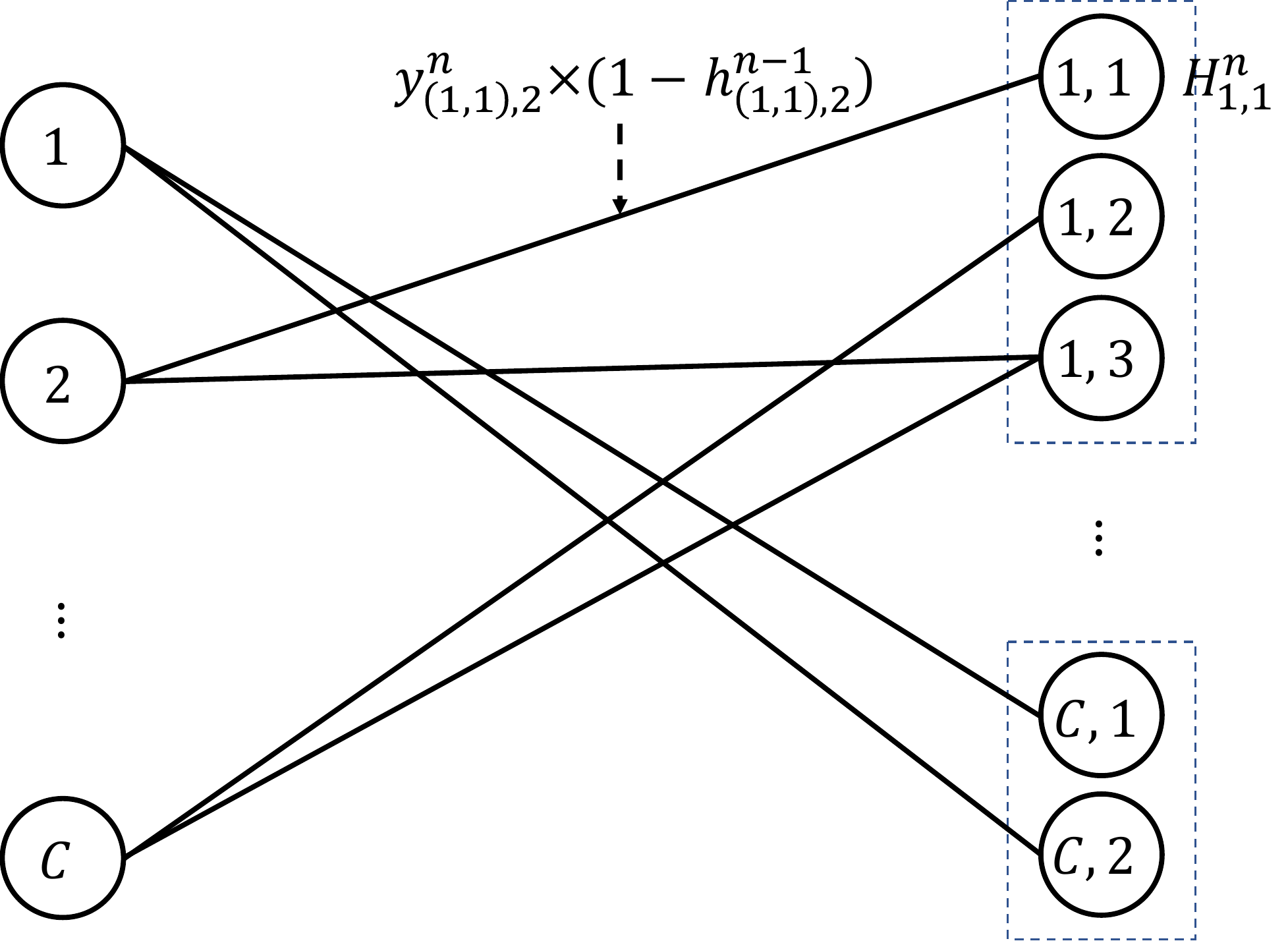}
\vspace{-4mm}
\caption{An illustration of $H^n_{i,k}$.}
\label{fig:Greedy}
\centering
\vspace{-3mm}
\end{figure}

Consider the bipartite graph shown in \figref{fig:Greedy} which has destination vehicles on the left and objects on vehicles on the right. The weight of each edge between a vehicle and an object corresponds to the reward if this vehicle receives that object. Then, $H^n_{i,k}$ can be computed by adding the weights of the edges connecting to the $(i,k)$ node.
%\kostas{This figure needs some editing: first, replace t with n-1 as the time step indicator in h/H and add n on y. second, since the nodes on the right are $(i,k)$ the last group of nodes correspond to $(C,*)$ and not to $(K,*)$. third I would have vertical dots to indicate multiple objects and vehicles. fourth I would not have the same number of objects per vehicle on the right to indicate they can be different}

\parae{Greedy max-weight scheduler.}
Motivated by this representation, we may use greedy solutions to the maximum weight matching problem of a bipartite graph \cite{MulticastMaxWeight}
to quickly find a good solution.
Specifically, at every frame/time step, the scheduler may select the transmission pairs based on decreasing order of the $H^n_{i,k}$ value, leading to the highest possible total weight/reward among the available transmissions for each frame, until there is nothing to deliver or the decision interval is over (i.e., $n = N$).

However,
the above weight does not take the size of an object, $s_{i,k}$, into account. Therefore, the scheduler may schedule an object which has a large weight but occupies a large portion of the frame, as opposed to scheduling a large number of smaller objects whose sum of weights might be larger than the weight of the single large object.
One way to address this is to 
divide the weight of an object by its size, and use the modified weight, 
$H^n_{i,k}/s_{i,k}$ instead, 
corresponding to a \textit{normalized reward} 
$y^n_{(i,k),j}/s_{i,k}$, over
the size of the object. 
% To avoid starving large objects, \sysname compensates for starvation in the following decisions (see details below). 
% \ramesh{Instead of more details later, add a section reference, or say "later in this section" or "below"}
%\kostas{hang and po han actually divided the weight in the implementation by the "transmission time" $s_{i,k}/B$ but in case of OFDMA the actual transmission time is the  transmission time of the whole frame and its best to avoid getting there.}
\vspace{-2mm}
\begin{algorithm}

	\caption{Greedy Max-Weight Scheduler}
	\begin{algorithmic}[1]
	\INPUT $y^n_{(i,k),j}$, $h^n_{(i,k),j}$, $s_{i,k}$, and $S^n_f$
	\OUTPUT $\pi(S^n) = \{x^n_{i,k}\}$
	\For {$n \in \{1,...,N\}$}
	\State Calculate $H^n_{i,k}$ from Eq. (\ref{eq:GreedyMetric}).
	\While {$\sum_{\forall i \in \mathcal{C}}\sum_{\forall k \in \mathcal{K}}{s_{i,k}\times x^n_{i,k}} \leq S^n_f$}
	\State Select a TX pair $(i,k)$ with the largest  $H^n_{i,k}/{s_{i,k}}$.
	\State Set $x^n_{i,k} = 1$.
	\EndWhile
	\State Update $h^n_{(i,k),j}$ based on the vehicle environment.
	\EndFor
	\end{algorithmic}
	\label{alg:Greedy}

\end{algorithm}
\vspace{-2mm}

We summarize in pseudo code the proposed greedy Max-Weight algorithm (Algorithm \ref{alg:Greedy}).
The time complexity of the scheduler can be easily shown to equal $\mathcal{O}(NCK\log(CK))$.
CPU experiments show that the Greedy Max-Weight scheduler runs fast and achieves near optimal performance for the scenarios of interest that we have studied (\secref{sec:eval}).

\parae{FPTAS-based scheduler.}
We also propose to use a well known fully-polynomial time approximation scheme (FPTAS) \cite{Book:Algo} to solve the selection problem at every time step.
We first introduce a dynamic programming framework which solves the following equation:
\small
\begin{align}
DP(\mathcal{C}\times\mathcal{K}, S^n_f) = &\max\{DP(\mathcal{C}\times\mathcal{K} \setminus (i,k), S^n_f), \nonumber \\
& H^n_{i,k} + DP(\mathcal{C}\times\mathcal{K}\setminus (i,k), S^n_f-s_{i,k})\}. \label{DP}
\end{align}
\normalsize
We then formulate the scheduling problem at each time step as a binary Knapsack problem, and solve it using FPTAS. While more efficient than dynamic programming, FPTAS still has high computational complexity (see \secref{sec:eval}).

\parab{Starvation compensation.} When network capacity is insufficient to transmit all objects, some objects may ``starve'' (not be transmitted). To avoid this, \sysname discards stale updates (point clouds), increases such objects' weight to maximize the likelihood that new updates (if still relevant with positive rewards) are transmitted in a future decision interval. Let $m$ represent the number of decision intervals over which an object has \textit{not} been transmitted even though it is still occluded and relevant. We replace in Algorithm 1 the value of $H^n_{i,k}/{s_{i,k}}$ with $\frac{H^n_{i,k}}{s_{i,k}/\sigma^2_{s_{i,k}}} \times m/\sigma^2_m$, where $\sigma^2$ denotes the variance and is used to make sure the contributions of the size and the starvation effects are normalized. 

\parab{Other details.} When a schedule is decided, the cooperative execution of that schedule among vehicles has differences depending on which V2V technology is used. DSRC uses TDMA among vehicles. LTE-V may use TDMA (mode 4) or SC-FDMA (mode 3), an OFDM variant, where frequency-time slots can be assigned to vehicles based on the schedule \cite{3GPP:36.885}. \sysname can employ DSRC or any LTE-V mode; in \secref{sec:eval}, we use real DSRC radios to demonstrate the scheduled transmissions, and have left LTE-V integration to future work.

%%%%%%%%%%%%%%%%%%%%%%%%%%%%%%%%%%%%%%%%%%

\section{Data Plane}
\label{s:data-plane}

Autonomous vehicles use 3D sensors for \textit{perception}, and a \textit{planning algorithm} to determine the vehicle's trajectory. \sysname  proposes to extend today's autonomous driving with \textit{cooperative perception}. Its data plane achieves cooperative perception using \textit{spatial reasoning} algorithms that generate object maps (\secref{sec:protocol}), and a \textit{planner} that relies on cooperative perception to improve driving safety.

% One of the key component of 3D perception is to reason the space about surrounding objects' position, dimension, and predict their motion, in order to plan a correct and safe trajectory and corresponding driving decisions.  In this section, we introduce these by describing two subcomponents in the data plane, spatial reasoning and trajectory planning, to demonstrate how \sysname can exploit perception to facilitate sensor sharing and leverage shared sensor to make better and safer driving decisions.

% Current technologies enhances the perception accuracy and robustness using a suite of multi-modality sensors (\eg  lidars, radars, cameras) and neural networks to understand these sensor inputs at the finest resolution possible. \sysname abstracts away the complication, uses lidar and occupancy analysis to demonstrate the effectiveness and benefits of sensor sharing.

\subsection{Spatial Reasoning} \label{sec:assess}

This component processes each frame of the LiDAR output and generates the object maps. Specifically, for each object $k$ in vehicle $i$'s view, spatial reasoning determines the \textit{visibility} $v_{(i,k),j}$ of that object with respect to another vehicle $j$, and the \textit{relevance} $r_{(i,k),j}$ of that object to that vehicle  (\secref{sec:protocol}). To do this, it must (a) detect roadway objects within its view, (b) assess their geometric and temporal relationships.

\parab{Extracting roadway objects}.
Several deep learning networks exist ~\cite{pointrcnn, voxelnet, pointpillars, centerpoint} that can detect objects in LiDAR frames. 
% and forecast motions~\cite{FastFurious}
However, these can be slow, require significant compute resources, are sometimes inaccurate, and generate information (\eg identify object classes and bounding boxes) that \sysname does not need. For \sysname, we simply need point clouds of stationary or moving objects on the road.

\begin{figure}
\centering
% \vspace{-2mm}
\includegraphics[width=0.6\columnwidth]{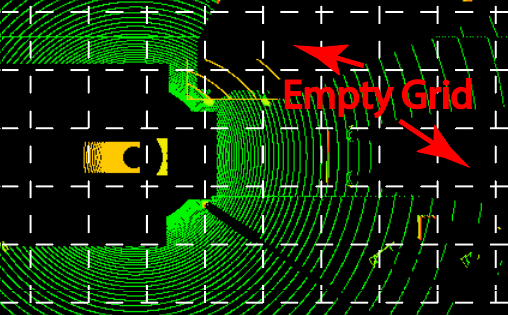}
\vspace{-2mm}
\caption{Empty occupancy grids indicate occluded area. }
\label{fig:occlusion}
% \vspace{-3mm}
\end{figure}

To extract these roadway objects, we voxelize~\cite{FastFurious, pointpillars} the point cloud by imposing a fine 2-D occupancy grid from the birds-eye-view perspective of the LiDAR point cloud (\figref{fig:occlusion}). More precisely, each 2-D grid element is a rectangular tube extending vertically on the z-axis.  Each point in the LiDAR frame falls into exactly one grid element. Each grid element may contain: (a) no points, (b) only points on the ground, or (c) points above ground. \sysname can determine if a point lies on the ground or above the ground because it knows the coordinates of the point, and the height of the LiDAR above the ground. Furthermore, \sysname assumes that each 2-D grid is labeled as either on the roadway surface or not. This information can be obtained by running a segmentation algorithm for drivable space detection on the 3-D map \cite{HDmapCVPR16, DeepSemantic}. Thus, all points in a 2-D grid element of type (c) which are above the ground constitute points belonging to a roadway object. The object itself consists of all contiguous type (c) grid elements (as we discuss later in \secref{sec:eval}, we use sub-meter grid dimensions so it is unlikely that points belonging to two different vehicles would fall into the same grid). Objects detected by different vehicles with the same grids (after perspective transformation at the receiver~(\secref{sec:planning})) are identified as the same unique object.

%\hangq{here is a bit tricky, the submeter grid may be fine for vehicles, but difficult for peds. }

% \sysname processes the sensor in data plane to analyze the space from each vehicle's perspective. To assess the relevance of each object to be shared for each recipient, we consider two factors: whether the the object is visible to the recipient, and whether the awareness have an impact on driving decisions.

\parab{Visibility determination.}
Having extracted all objects in its view, to determine $v_{(i,k),j}$, vehicle $i$ simply traces a ray from vehicle $j$'s current position around every object $o_{i,k'}$ in its own view ($o_{i,k'}$ includes the vehicle $i$ itself, if visible). If $o_{i,k}$ falls into the shadow of $o_{i,k'}$, then the latter object occludes the former and $v_{(i,k),j}$ is false. If no such $o_{i,k'}$ exists, and if $o_{i,k}$ is within the LiDAR range of $j$, then $v_{(i,k),j}$ is true. %\ramesh{Need a figure which shows the two occlusion cases, one by the car itself and one by another object seen by the car.}

\parab{Relevance determination.}
The intuition behind relevance determination is that some objects may not be  relevant to other vehicles, even if invisible to those vehicles. For instance, if a vehicle is turning right at an intersection,  it is unlikely to need information about vehicles driving straight on the opposite lane. In \sysname's  implementation, an object is relevant to another vehicle if the trajectories of those two objects could potentially collide at some point in the future. 

More precisely, $r_{(i,k),j}$ is a value that assesses whether vehicle $j$'s trajectory can collide with $o_{i,k}$. Vehicle $i$ gets $j$'s trajectory from control messages. It obtains $o_{i,k}$'s trajectory by estimating this objects' heading and velocity continuously over successive frames.  By extrapolating these trajectories, \sysname can determine if the two trajectories collide at some point in time. Given this, one can define $r_{(i,k),j}$ in two ways: (a) as a boolean value that is true when $j$ can collide with $o_{i,k}$, or (b) as the reciprocal of the time to collision (a value between 0 and 1, assuming time is in milliseconds). The intuition for the latter choice is clear: objects that $j$ is likely to encounter sooner are more relevant\footnote{At large scale, boolean and reciprocal definitions perform similarly  (\eqnref{eq:reward_func}). For simplicity, we used boolean in the evaluation.}.

\parab{Loss compensation.} Similar to handling control message losses (\secref{sec:protocol}), data packet losses are also compensated by extrapolation. Given a loss of a particular object, \sysname calculates the center of the object point cloud using the heading and velocity estimated from the location and timestamp of previous receptions. Then, it translates the last received point cloud to the newly estimated center location.

\subsection{Trajectory Planning}
\label{sec:planning}

Autonomous vehicles use sensor inputs to make driving decisions. These driving decisions occur at three different scales: \textit{route planning} occurs at the granularity of a trip, \textit{path and trajectory planning} occurs at the granularity of a road segment (a few tens to hundreds of meters), and \textit{low-level control} ensures that the vehicle follows the planned trajectory by effecting steering and speed control. In \sysname, vehicles must make these decisions, by incorporating received point-clouds into their own LiDAR output.

In this paper, in order to  quantify the end benefits of  cooperative perception, we develop a path and trajectory planning algorithm that incorporates objects received from other vehicles. The large, existing literature on this topic (see, for example,~\cite{UCBMotionPlanningSurvey, RRT,FMT}) does not take cooperative perception into account. Recent research has recognized and incorporated partial visibility into trajectory planning~\cite{PavoneMapPredict,MPNet}, but relies on training data and  predictions of the geometry of the invisible area. In contrast, we develop a planning algorithm using concrete cooperative perception for trajectory planning.

\parab{Perspective transformation.}
Before it can plan a trajectory, \sysname must \textit{re-position} the received point clouds into its own LiDAR output. It uses the 3-D map for this. The sending vehicle positions the point cloud in its own coordinate frame of reference. To re-position it to the receiver, let \(T_{s}\) be the transformation matrix from the sender's coordinate frame of reference to that of the 3-D map and \(T_{r}\) be the transformation matrix for the receiver. To transform a point \(V_s\) in the sender's view to a point \(V_r\) in the receiver's, \sysname uses: \(V_r = T_{r}^{-1} * T_{s} * V_s\). Having done this, it updates each occupancy grid (\secref{sec:assess}) with the received point cloud, then uses the occupancy grid to determine a \textit{path} and then a \textit{trajectory}.

% To provide visibility and spatial reasoning into the occluded areas, \sysname can augment each vehicle's spatial reasoning results by merging the shared sensor into their own perspectives. Transforming the perspective of the sensor requires the relative pose of the sender and receiver. Current technologies use  simultaneous localization and mapping (SLAM) \cite{loam, vloam} to obtain the pose of autonomous vehicles with respect to an HDMap. Recent literature~\cite{CarMap} also focus on how to collect and update such a map for more accurate positioning. \sysname assumes the availability of such a map and computes a transformation matrix $T$ based on vehicle pose. By definition, any point (\(V = [x,y,z]\)) in the vehicle domain can be transformed to a point in the map domain (\(V' = [x',y',z']^T\)), where
% \([x,y,z,1]^T = T * [x,y,z,1]^T\). Now, suppose that the sender's
% transformation matrix is \(T_{s}\) and the receiver's is \(T_{r}\),
% then, the receiver can compute the perspective transformation of a point
% \(V_s\) in the sender's view to a point \(V_r\) in the receiver's view
% as follows: \(V_r = T_{r}^{-1} * T_{s} * V_s\). Placing $V_r$ in the occupancy grid, the occluded objects can now replace the empty grid, which can affect the path planning results in the next step.

\parab{Path Planning.} This step determines a viable and safe path through drivable space that avoids all objects. It uses the occupancy grid defined above (\secref{sec:assess}) after augmenting it with received objects. To understand path viability, recall that each grid element can either have one or more points belonging to an object, or be \textit{unoccupied}. Moreover, using the 3-D map, we can annotate whether a grid element belongs to \textit{drivable} space or not, and also whether a vehicle can traverse the grid element in both directions or uni-directionally.

The input to path planning is a source grid element and a target grid element.
The output of path planning is a \textit{path} in the 2-d grid, where the width of the path is the width of the car, and every grid element that intersects with the path must (a) be unoccupied and (b) be drivable in the direction from the source to the target. %\ramesh{Need a figure to show this.}
\sysname uses $A\star$ heuristic search~\cite{Astar} to determine a valid path. We constrain the search so that the resulting path is \textit{smooth}: \ie it does not have sharp turns that could not be safely executed at the current speed.

\parab{Trajectory Planning and Collision Avoidance.} On the resulting path, \sysname picks equally spaced \textit{waypoints}; a \textit{trajectory} is a collection of waypoints and associated times at which the vehicle reaches those waypoints. To determine those times, the trajectory planner must determine a \textit{collision-free} trajectory; when the vehicle is at a particular waypoint, all other vehicles must be far enough from that waypoint. To determine this, \sysname uses the estimated trajectory of received objects, as well as estimates of the trajectory of vehicles within its own sensor's view. \sysname also calculates the time of arrival to and departure from this waypoint based on estimated speed and vehicle dimension. When a predicted collision is far enough, \sysname follows the planned trajectory until within stopping distance (based on current speed and brake deceleration) of that waypoint of collision. 

%\ramesh{Illustrate this in the same figure as the path.}

% To predict potential future collisions, \sysname first computes an estimated arrival time for each grid on the planned trajectory based on its current velocity and acceleration. For detected dynamic objects, \sysname also estimates the motion of the cluster of occupied grid based on the tracked velocity. When the planned trajectory collides or overlaps with the extrapolated trajectory of detect objects, \sysname check on the time domain whether a potential collision can be avoided. Even if it cannot be avoided, the trajectory is still valid if the predicted collision point is far away from current position. \sysname performs an emergency brake based on the distance and speed of the vehicle only when if the collision point is within a braking range of 2 seconds \cite{2secondrule} or a light break (<30\%) cannot prevent a collision. 

\section{Performance Evaluation}
\label{sec:eval}
\begin{figure*}
\centering
%\vspace{-1mm}
\includegraphics[width=\textwidth]{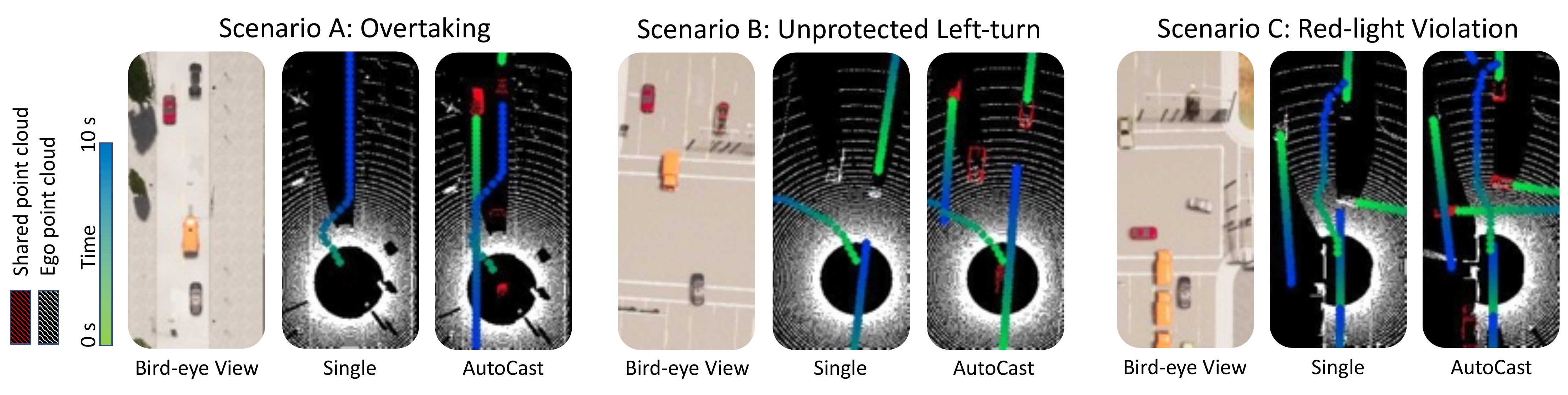}
\vspace{-8mm}
\caption{End-to-end evaluation scenarios: overtaking, unprotected left-turn, and red-light violation. A planner on the \textit{ego} vehicle (gray, bottom of the bird-eye view) finds a trajectory to navigate through each scenario without collision. The gradient trajectory color (green to blue) indicates a temporal horizon (closer to farther). The LiDAR views show the perception results using either non-sharing baseline (\textit{Single}), or \sysname. The red points in the LiDAR view are shared points, while the white ones are from the ego vehicle itself. In each scenario, a \textit{passive} (without  communication capability)  \textit{collider} vehicle (red), occluded by a truck (orange) and thus invisible from the ego's \textit{Single} view, 
% either coming from the opposite direction, or violating a traffic light, 
may cause a hazardous situation. \sysname makes the ego vehicle aware of the collider so the ego can react early to avoid a collision.}
\label{fig:scenes}
\vspace{-5mm}
\end{figure*}

In this section, we first evaluate \sysname end-to-end: we show that cooperative perception can improve driving safety on three autonomous driving benchmarks (\secref{s:end-end-scenario}). We breaks down the results by traffic density to discuss scalability (\secref{sec:scalability_eval}), and then details the results of each scenario (\secref{sec:per_scene_eval}). Next, we evaluate the latency of each processing module (\secref{sec:microbenchmark}), and validate transmissions using  real V2V radios (\secref{sec:dsrc_exps}). We conclude with micro-benchmarks of scheduling details (\secref{sec:scheduler_eval}).

% In this section, we evaluate \sysname end-to-end under realistic environments.
%

\subsection{Methodology}
\label{s:methodology}

\parab{The Carla simulator.}
% Evaluating autonomous driving in general, and \sysname in particular, poses significant challenges. To address this, industry and academia have recently developed photo-realistic simulation platforms like Carla~\cite{CarLA}. 
Carla~\cite{CarLA}, a photo-realistic simulation platform, uses a game engine to simulate the behavior of realistic environments, and contains built in models of freeways, suburban roads, and downtown streets. Users can create vehicles that traverse these environments and attach advanced sensors such as LiDAR, Camera, Depth Sensor to them. As these vehicles move through the environment, Carla simulates environment capture using these advanced sensors. Users can design planning and control algorithms using the captured environment to validate autonomous driving.

\parab{Implementation.}
We have implemented the scheduler, spatial reasoning, and trajectory planning in Carla. The total \sysname implementation is 27,124 lines of code. In addition, to configure the scenarios, we have developed on top the Carla autonomous driving challenge \cite{carlaadchallenge} evaluation code. In our implementation, all vehicles use Carla's default longitudinal and lateral PID controller as the lower-level control to steer the vehicle along the planned trajectory.

To simulate metadata exchange between vehicles, we have incorporated V2V. Specifically, our implementation models LTE-Direct QPSK with 10 MHz bandwidth~\cite{LTE-Direct}, which translates to a peak rate of $\sim$ 7.2~Mbps. We implement LTE-Direct TDMA Mode 4~\cite{LTE-Direct} and simulate V2V channel loss in all scenarios using models described in 3GPP standards \cite{3GPP:36.785,3GPP:36.885}.

\parab{End-to-end evaluation scenarios.} To demonstrate the benefits of \sysname end-to-end, we have implemented three scenarios (\figref{fig:scenes}) from the US National Highway Transportation Safety Administration (NHTSA) Precrash typology~\cite{NHTSA-precrash}. In these, occlusions can impact driving decisions. 
\begin{description}
\vspace{-2mm}
\item[Overtaking] A stopped truck on a single-lane road forces a car to move to the lane with on-coming traffic. The truck occludes the car’s view of the opposite lane. 
\vspace{-2mm}
\item[Unprotected left turn] A car and a truck wait to turn left in opposite directions at an intersection. The truck blocks the car’s view oncoming traffic.
\vspace{-2mm}
\item[Red-light violation] A truck waits to turn left at an intersection, and a car drives straight towards the intersection. Another car jumps the red-light in the perpendicular direction; the violator is occluded by the truck.
% and invisible to the car crossing the intersection.
% the truck can see the violator, but the car crossing the intersection cannot see the violator.
\vspace{-2mm}
\end{description}

% \parab{Scenarios.} \figref{fig:scen6,fig:scen8,fig:scen10} shows the three scenarios we have implemented for evaluation: \textit{triangle overtake}, \textit{unprotected left turn}, and \textit{red light violator}. In triangle overtake, a parking truck is blocking the way of a sedan in a double-way single lane road with a dashed yellow lane divider. The truck is also occluding the sedan’s view of the opposite lane. An autonomous agent has to make a lane change maneuver to overtake. In unprotected left turn, a left-turning sedan encounters another left-turning truck in the opposite direction at an intersection. The truck is blocking the sedan’s view of the opposite lane and potential straight-going vehicles. An autonomous agent has to make the unprotected left turn without collision. In red light violator, a left-turning truck and a straight-going sedan is trying to cross the intersection under green light, when another sedan rushes a yellow light and/or violates a red light on the perpendicular way. The truck can see the violator and stops to avoid collision. But it is blocking the sedan’s view. An autonomous agent has to cross the intersection without collision. In all three scenarios, we apply the proposed spatial reasoning and trajectory planning and use Carla's default longitudinal and lateral PID controller as the lower-level control to steer the vehicle along the planned trajectory. \figref{scen6_pointcloud, scen8_pointcloud, scen10_pointcloud} shows, in each scenario, the birds-eye-view of the point cloud and the planned trajectory with and without sensor sharing.

\parab{Experiments with real radios.} To demonstrate that an implementation of \sysname can plausibly work over real radios, we run \sysname on a small-scale testbed using three iSmartWays DSRC radios~\cite{ismart}. In these experiments, we record the trace data from all scenarios; for each frame (every 100ms), the trace includes all exchanged metadata, the computed schedule, and the object point clouds. We then playback the trace over DSRC radios to validate if the scheduled transmissions complete in time.

\parab{Baselines.}
We compare \sysname against an approach in which each car makes trajectory planning decisions based on its own sensor alone (called \textit{Single}), and one in which cars within range deliver objects in a round-robin fashion (called \textit{Agnostic}). We also implemented EMP~\cite{emp}, an edge-assisted cooperative perception scheme, as another baseline. Because EMP is a V2I solution instead of V2V, for fair comparisons, we assign EMP a total bandwidth of 50~Mbps, but use 7.2~Mbps for Agnostic and \sysname. In quantifying the efficacy of our greedy scheduler, we also compare it with 1) \textit{FPTAS}, 2) Optimal using dynamic programming.

\parab{Metrics.}
We use several metrics to evaluate \sysname. In end-to-end experiments, we quantify scenarios \textit{outcomes} (\eg a crash, or a near miss), the \textit{reaction} time (between when a vehicle detects a potential collision  and the time needed for it to avoid the collision), and the \textit{closest distance} between two vehicles at any point during the scenario. 
To analyze the scalability, we compare the \textit{collider visibility} in terms of the number of visible frames and shared points under different schemes at different traffic densities.
To evaluate the scheduler's efficacy, we quantify
\textit{rewards}, 
% and the \textit{reward ratio} (the percentage of the object reward is received), 
\textit{time complexity}, and \textit{scheduled delay} of objects with different rewards.

\begin{figure}
\centering
    
    \includegraphics[width=0.8\columnwidth]{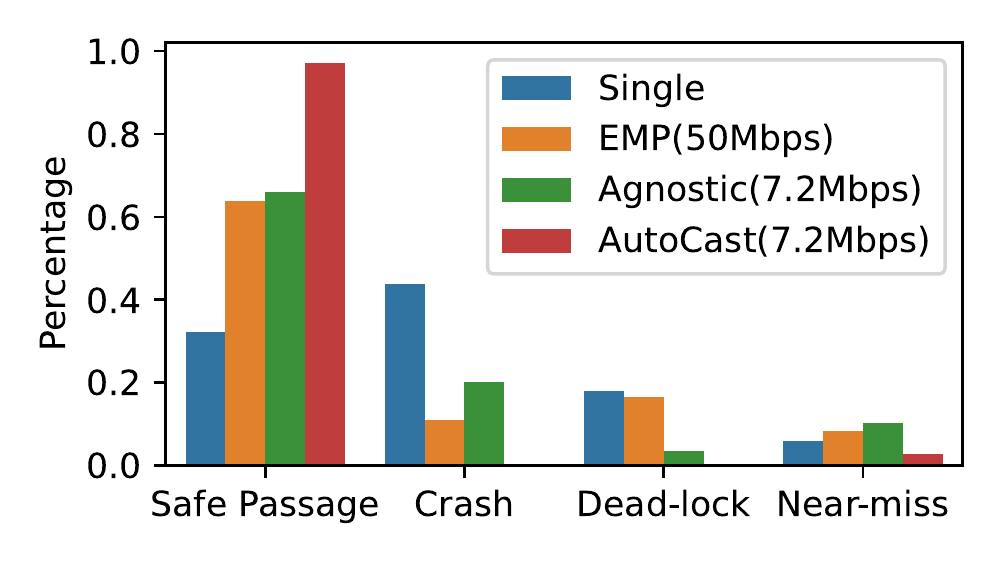}
    \vspace{-5mm}
    \caption{Scenario Outcome}
    % \vspace{-3mm}
    \label{fig:scalability_performance}

\end{figure}

% \begin{figure}
% \centering
% %\vspace{-1mm}
% \includegraphics[width=0.9\columnwidth]{fig/scalability/safepass_vs_density.eps}
% \vspace{-1mm}
% \caption{Safe Passage vs. Traffic Density}
% \label{fig:safepass_vs_density}
% \vspace{-1mm}
% \end{figure}

\begin{figure*}
\centering
\begin{minipage}{0.73\columnwidth}
\centering
\includegraphics[width=0.85\columnwidth]{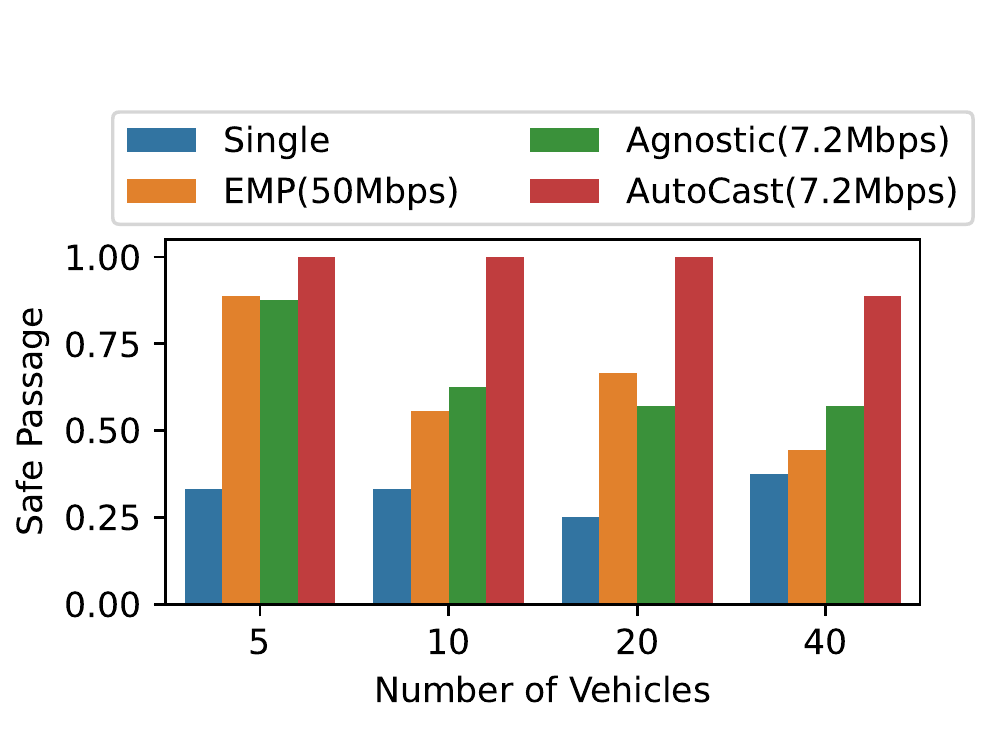}
\vspace{-4mm}
\caption{Safe passage vs. traffic density}
\label{fig:safepass_vs_density}
\vspace{-5mm}
\end{minipage}
\hfill
\begin{minipage}{1.33\columnwidth}
\centering
\includegraphics[width=0.48\columnwidth]{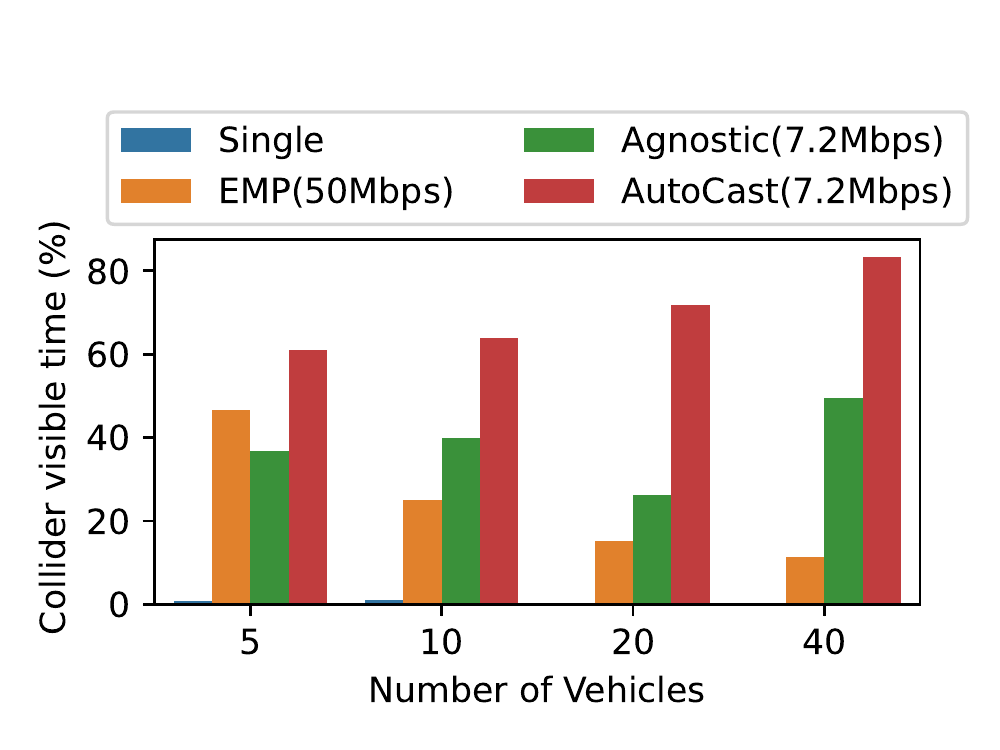}
\includegraphics[width=0.48\columnwidth]{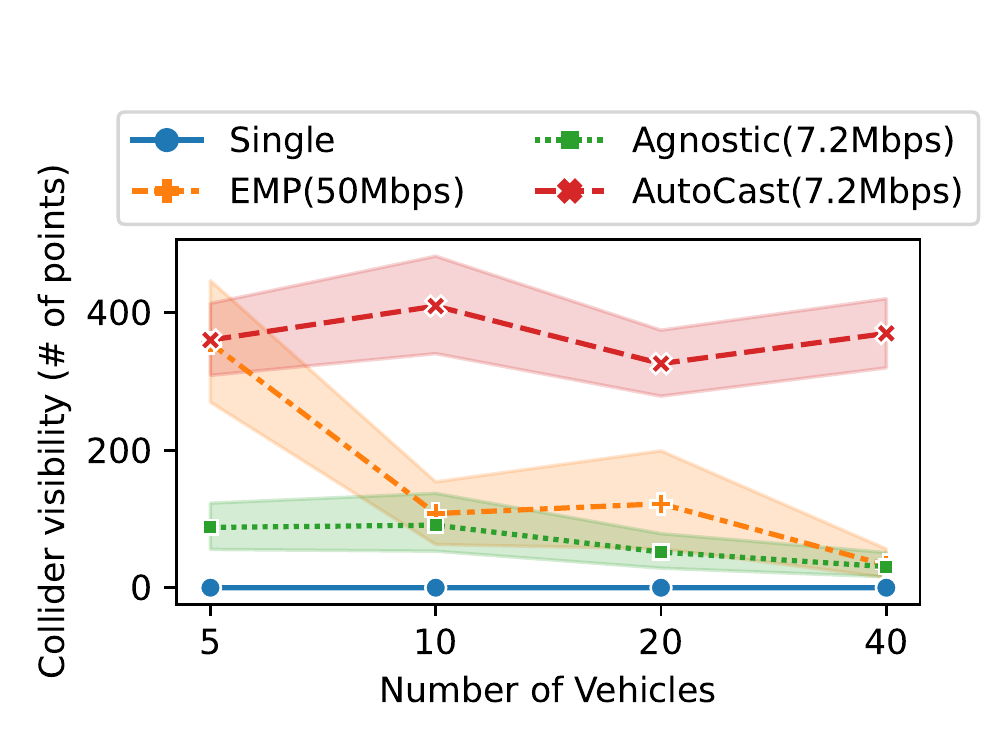}
\vspace{-4mm}
\caption{Collider visibility: visible time and visible size vs. traffic density}
\label{fig:cubics_vs_density}
\vspace{-5mm}
\centering
\end{minipage}
% \hfill
% \begin{minipage}{0.66\columnwidth}
% \centering
% \includegraphics[width=0.9\columnwidth]{fig/scalability/time_vs_density.pdf}
% \caption{Collider visible time vs. traffic density}
% \label{fig:time_vs_density}
% \centering
% \end{minipage}
\end{figure*}
\subsection{End-to-end Scenario Evaluation}
\label{s:end-end-scenario}

\parab{Goal.}
The NHTSA pre-crash typology defines a set of challenging scenarios. In this section we seek to understand whether cooperative perception can result in safer driving outcomes than a system without this capability. We evaluate these scenarios in Carla: for each scenario, we explore different points of the scenario parameter space (described below), and record the metrics described above.

% In addition to showing a bird-eye view of each scneario, \figref{fig:scen6}, \figref{fig:Scen8}, \figref{fig:Scen10} also shows (on the right side of each figure) the point cloud of a single vehicle, and the merged point cloud from \sysname. On the point cloud, the figure also shows the planned trajectory and the estimated trajectory of other \textit{visible} vehicles. In all three scenarios, \sysname enables the awareness of an vehicle that is on the trajectory to collision which is not previous visible. This awareness triggers a very different control decisions. To systematically demonstrate the impact of the awareness on control decisions, we evaluate each scenario with a set of different configurations. 

\parab{Terminology and Experiment Setting.}
In each of our scenarios (\figref{fig:scenes}), there are three entities: the gray sedan is the \textit{ego vehicle} on which \sysname runs, the red sedan is a \textit{passive collider} which cannot communicate and only uses its own sensors to plan its trajectory, and the orange truck is an \textit{occluder}. In each scenario, the paths between the  ego and the collider intersect. We set up their speeds such that their \textit{trajectories} almost collide (\ie both vehicles would come very close to each other if both did not see each other at all).

Specifically, we generate several \textit{configurations} as follows. We set the \textit{base speed} of the collider to 3 different values (20~km/h, 30~km/h and 40~km/h). At a given base speed, the collider's trajectory would (in the absence of avoidance) intersect with that of the ego. A second dimension of the configuration is an \textit{intersection delta}; ranging from -2~s to +2~s (with steps of 0.25~s), a value of $\delta$ means the collider actually arrives at the intersection point $\delta$~s before (or after) the ego vehicle. This latter parameter controls how closely the two cars approach each other. This gives us a total of 24 different configurations for each scenario. For each configuration, we also vary the number of  vehicles (from 5 to 40 within range R (\secref{sec:protocol})) to evaluate the performance at different scales.

We present three sets of results. First, we present the end-to-end results for all scenarios that highlights \sysname's performance against \textit{Single}, \textit{EMP}, and \textit{Agnostic}.Then, we break it down by traffic density to highlight the scalability (\secref{sec:scalability_eval}) against the baselines. Finally, to illustrate subtleties in \sysname's design, we break down results by scenario (\secref{sec:per_scene_eval}), in a sparser traffic setting.

% For each scenario, we configure them so that ego vehicle and the collider will collide if each of them is on a dummy autopilot: no collision avoidance but only follow the default route. On top of this, we vary the speed of the colliding vehicle so that we can change the colliding vehicle's arrival time to the collision point. The range of this sweeping variation is set so that at lowest collider's speed, the ego vehicle would pass the collision point before the collider does without slowing down, and at highest collider's speed, the ego vehicle would also pass the collision point, but after the collider, without slowing down. 

% \hangq{change figure with bg traffic?}

% \begin{figure}
% \centering
% \centering
% \includegraphics[width=0.8\columnwidth]{fig/Scenario/crash_v2.eps}
% % \vspace{-3mm}
% \caption{Crash and Near Miss.}
% % \vspace{-3mm}
% \label{fig:crash}
% \end{figure}

\parab{Outcomes.}
With this setting, there are four possible outcome: \textit{safe passage}, \textit{near-miss}, \textit{crash}, and \textit{deadlock}. A near miss occurs when the ego and collider pass within 2~m of each other. In deadlock, which occurs only in Overtaking, both vehicles stop without colliding but neither can make forward progress. This situation is not inherently unsafe, it does represent an undesirable driving outcome where participants must coordinate to resolve the deadlock. Beyond these outcomes, we are also interested in quantifying the closest distance between vehicles during any point in the simulation, and the reaction time (the time between when the ego is aware of the collider to the last possible moment before it can start braking). %\figref{fig:crash} shows the outcomes across all scenarios and aggregated across all configurations.

% \begin{figure}
%     \centering
%     \includegraphics[width=0.8\columnwidth]{fig/Scenario/legend_2col.png}
    
%     \includegraphics[width=\columnwidth]{fig/scalability/crashdetail_density.eps}
%     \caption{Outcome Details vs. Traffic Density}
%     \label{fig:scalability_performance}
% \end{figure}
% \begin{figure}
%     \centering
%     % \includegraphics[width=\columnwidth]{fig/scalability/completionTime.eps}
%     \includegraphics[width=0.7\columnwidth]{fig/scalability/nearmiss_density.eps}
%     \caption{Closest Distance vs. Traffic Density}
%     \label{fig:nearmiss_density}
% \end{figure}

\parab{Results: Dense Traffic.} Our first set of results demonstrate the efficacy of \sysname in settings that it was designed for: highly dense settings with a large number of traffic participants, where channel capacity precludes transmission of point clouds of all participants.
%
% \parae{Methodology.}
In this experiment, we ran all three scenarios (overtaking, left turn, red light violation), but varied the traffic density from 5 to 40 vehicles within range R (\secref{sec:protocol}). For each scenario, we swept the same configuration dimensions with collider speed varying from 20-40 km/h.

% \hangq{revisit, include EMP, maybe cut down closest distance, check maybe revise density}
% \parae{Results: Outcomes.}
\figref{fig:scalability_performance} plots the fraction of outcomes for each of the four alternatives we consider (Single, EMP, Agnostic, and \sysname). \sysname ensures safe passage at all collider speeds in all scenarios. By contrast, Single, which does not use cooperative perception, incurs crashes about half the time, many deadlocks and some near misses. 
EMP~\cite{emp} divides full point clouds into segments of a voronoi diagram. Each vehicle transmits the closest segment to an edge server, which then can forward to other vehicle recipients. Using an order of magnitude higher V2I communication bandwidth, EMP safely passes half of the traces, incurs about 20\% crashes and 20\% deadlocks, most of which happens at high density scenarios. EMP does not consider object relevance, or prioritize segments, so suffers from lower collider visibility (see scalability results in \secref{sec:scalability_eval}).
Agnostic, which also does not prioritize transmissions, but benefits from object-based transmission, passes 20\% more traces with lower V2V bandwidth. Similar to EMP, at higher density, Agnostic exhibits an undesirable outcome in about a third of scenario settings. 
% (see \secref{sec:scalability_eval}). 
% Single, on the other hand is safe for only about 40\% of the scenario settings. In contrast to these two, \sysname is near-perfect: at the highest traffic density we have evaluated, it incurs a small number of near misses.

% \parae{Results: Closest Distance.}
We discuss differences between \sysname, EMP and Agnostic below. \sysname outperforms EMP and Agnostic for two reasons: (a) it extracts objects and prioritizes on cheap transmissions based on visibility and relevance, and, (b) when data to be transmitted exceeds the channel capacity, and some objects have to be left out, \sysname compensates by prioritizing these objects in the next decision interval. This ensures consistent and continuous updates for critical invisible objects. In EMP and Agnostic, due to delayed or missing updates, the closest distance to the collider is often below 2 m, leaving the ego vehicle almost 0 reaction time. 

Next, we show more details by analyzing why \sysname can outperform other baselines at scale (\secref{sec:scalability_eval}), and breaking down the results by scenario (\secref{sec:per_scene_eval}) for detailed analysis.

\subsection{Scalability Results}
\label{sec:scalability_eval}
We now show the scalability of \sysname by breaking down the evaluation results (\figref{fig:scalability_performance}) by traffic density. 

\parab{Safe passage at different traffic densities.}
\figref{fig:safepass_vs_density} shows the percentage of safe passage of all scenarios with the number of vehicles varying from 5 to 40. 
The \textit{Single} baseline shows a uniform performance across traffic densities: using ego vehicle's LiDAR alone can only pass around 25\% of the traces. It is interesting to see at very high density (40 vehicles), the rate is slightly higher because the road becomes crowded,  forcing all participants, including the ego vehicle, to slow down and proceed with caution.  
In contrast, EMP and Agnostic achieve on average around 60\% safe passages (\figref{fig:scalability_performance}), but most success concentrate at low density. As the traffic density increases, it is getting harder to avoid collisions and near-misses. We discuss key insights into poor scalability of both approaches.
Given an order of magnitude higher bandwidth for V2I, EMP is able to share almost all Voronoi segments~\cite{emp} at low density to cover the entire area. However, as the number of vehicles increases, each vehicle is transmitting a smaller and smaller segment around it, where the point cloud is the densest. Therefore, the total number of points to share, covering the same area, increases and exceeds even the V2I bandwidth limit. Also, since EMP does not prioritize segments based on relevance to receivers, the ego vehicle is aware of the collider only when the particular segment, where the collider is in, gets transmitted.
Agnostic, which isolates objects to reduce bandwidth requirement, but does not prioritize object transmission,  incurs a few near misses at low densities. As density increases, Agnostic incurs more crashes and deadlocks; with 10 vehicles and above, the probability of collider being transmitted is very low; Agnostic exhibits an undesirable outcome in about half of scenario settings. 
Over all traffic densities, \sysname is near-perfect: it maintains 100\% safe passage to 20 vehicles; at the highest traffic density we have evaluated, it incurs a small number of near misses without any collisions. Using only 7.2~Mbps, \sysname scales gracefully to up to 40 vehicles in a distributed fashion, not depending on infrastructure support.

% \begin{figure*}
% \centering

% \begin{minipage}{0.66\columnwidth}
%     \centering
    
%     % \includegraphics[width=0.6\columnwidth]{fig/Scenario/legend_2col.png}
%     \includegraphics[width=\columnwidth]{fig/scalability/outcome_ratio.pdf}
%     \vspace{-5mm}
%     \caption{Scenario Outcome}
%     \vspace{-3mm}
%     \label{fig:scalability_performance}
% \end{minipage}
% \hfill
% \begin{minipage}{0.66\columnwidth}
% % \vspace{2mm}
%     \centering
%     % \includegraphics[width=\columnwidth]{fig/scalability/completionTime.eps}
%     % \includegraphics[width=0.9\columnwidth]{fig/scalability/nearmiss_density.eps}
%     \includegraphics[width=0.9\columnwidth]{fig/scalability/nearmiss_density-eps-converted-to.pdf}
%     % \vspace{-1mm}
%     \caption{Closest distance}
%     \vspace{-5mm}
%     \label{fig:nearmiss_density}
% \end{minipage}
% \hfill
% \begin{minipage}{0.66\columnwidth}
% \centering
% % \includegraphics[width=\columnwidth]{fig/Scenario/crash_v2.eps}
% \includegraphics[width=\columnwidth]{fig/Scenario/crash_v2-eps-converted-to.pdf}
% \vspace{-5mm}
% \caption{Per-scenario at low-density}
% \vspace{-5mm}
% \label{fig:crash}
% \end{minipage}

% \end{figure*}

\begin{figure}
\centering
\includegraphics[width=0.66\columnwidth]{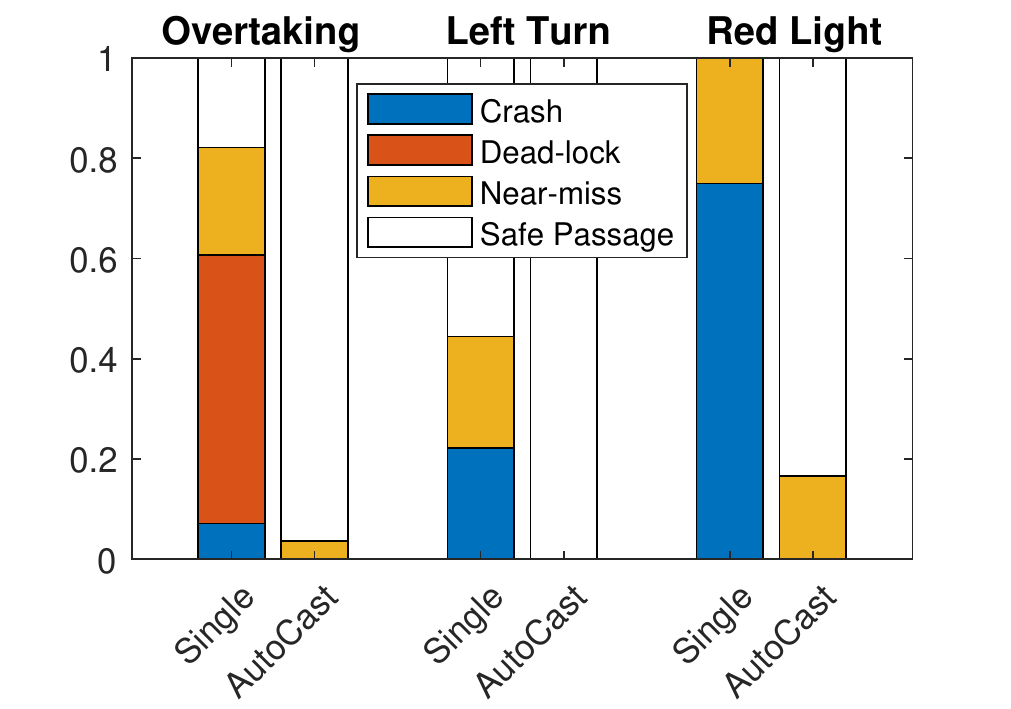}
\vspace{-5mm}
\caption{Per-scenario outcome at low-density}
% \vspace{-3mm}
\label{fig:crash}
\end{figure}

\parab{Collider visibility.}
To better understand the scalability results, we pick one intersection scenario (Scenario C: red-light violation) to evaluate and compare the \textit{collider visibility} against different baselines. 
\figref{fig:cubics_vs_density} shows the percentage of frames where collider is visible (left) and the size of the shared point cloud of the collider (right). Because both Agnostic and EMP does not prioritize transmissions, the chance of the collider being transmitted is highly dependent on the number of objects (or segments in EMP's case) to be transmitted and their sizes: at low traffic density, EMP is likely to transmit all segments, Agnostic has a higher chance to transmits the collider among less other objects; at high density, EMP cannot transmit all segments, and Agnostic renders lower collider visible time as well. 
It is critical for the planner to have high and stable collider visibility, because how often and how big the point cloud of the collider gets transmitted directly determines the object detection accuracy, reaction time, the trajectory planning, the control decisions, and therefore the scenario outcome (see per-scenario analysis \secref{sec:per_scene_eval}).  \sysname enables safe passage at high traffic density by providing consistent and stable (>80\% of the time) visibility into occluded and relevant objects to avoid safety hazards.

\begin{figure}
\centering
\centering
\makebox[\columnwidth][c]{
\includegraphics[width=\columnwidth]{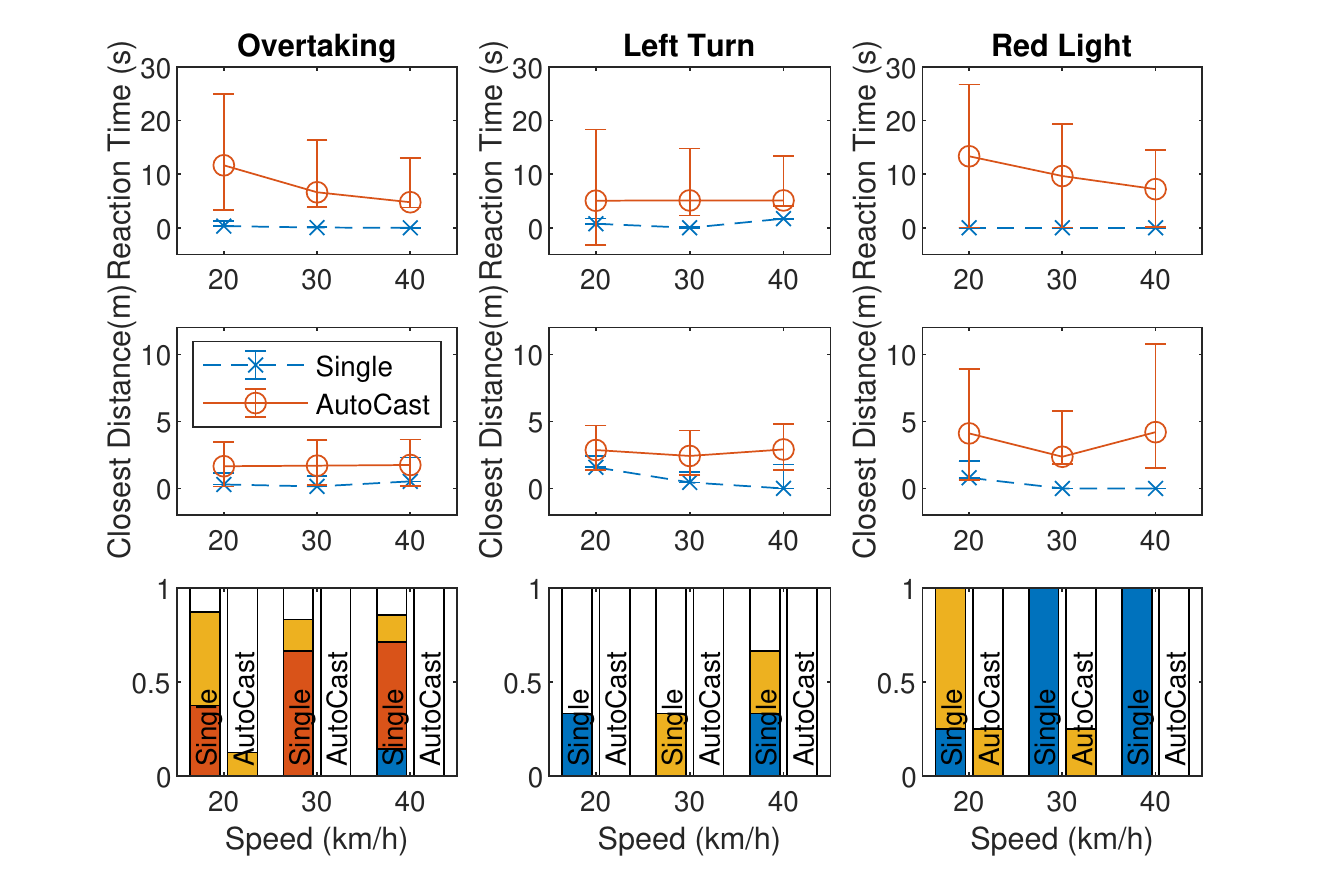}
}
\includegraphics[width=0.8\columnwidth]{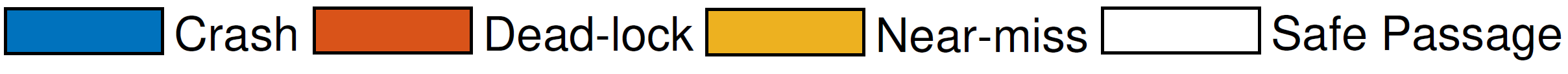}
\vspace{-3mm}
\caption{Reaction time, near-miss and crash details}
\vspace{-1mm}
\label{fig:crash_detail}
\centering
\end{figure}

\begin{figure*}
\centering
\begin{minipage}{0.9\columnwidth}
\centering
\includegraphics[width=\columnwidth]{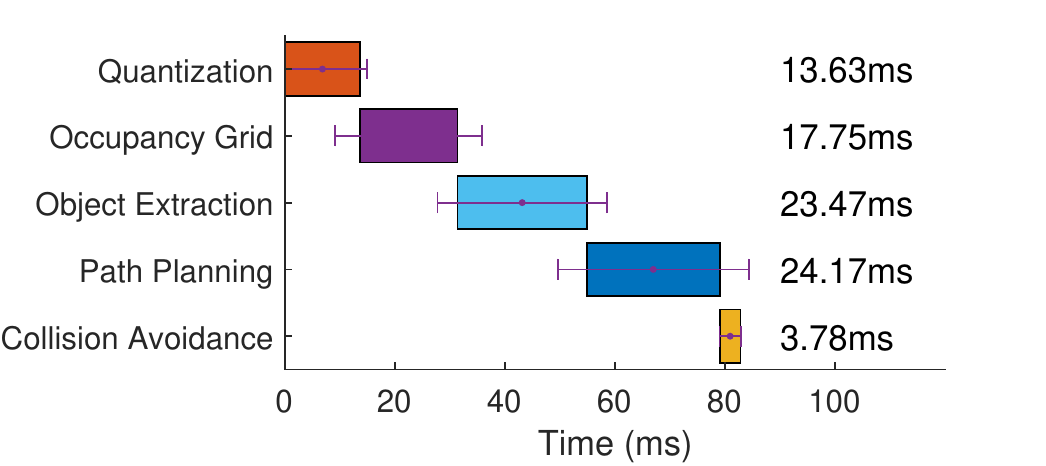}
\vspace{-5mm}
\caption{Pipeline micro benchmark}
\vspace{-5mm}
\label{fig:pipelineeval}
\end{minipage}
\hfill
\begin{minipage}{0.58\columnwidth}
\centering
\includegraphics[width=\columnwidth]{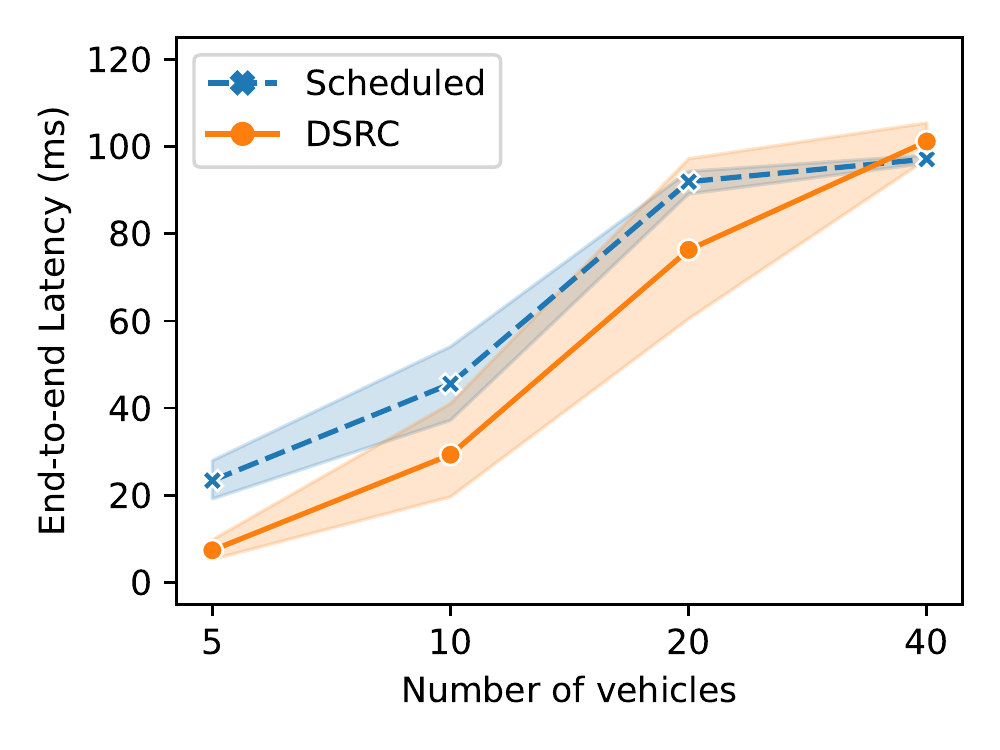}
\vspace{-7mm}
\caption{End-to-end latency per decision interval.}
\label{fig:dsrc_latency}
\vspace{-7mm}
\end{minipage}
\hfill
\begin{minipage}{0.58\columnwidth}
\centering
\includegraphics[width=\columnwidth]{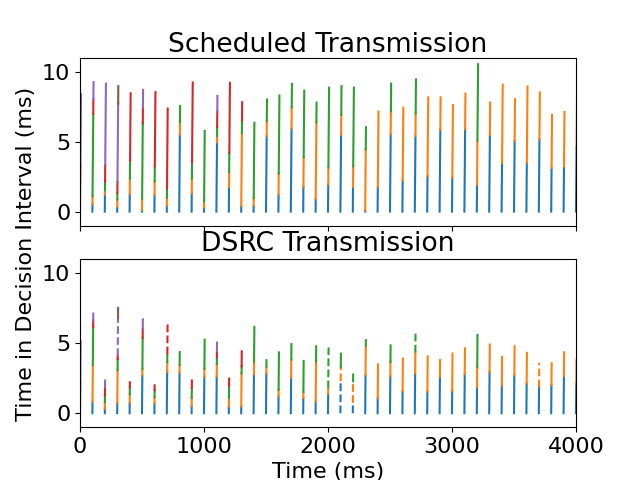}
\vspace{-4mm}
\caption{Transmissions in the overtaking scenario at low density}
\vspace{-5mm}
\label{fig:scen6_dsrc}
\end{minipage}

\end{figure*}

\subsection{Per-scenario Results}
\label{sec:per_scene_eval}
We now illustrate the benefit of cooperative perception in each of these three scenarios. We do this at low density, because the traffic dynamics are simpler and the results are easier to understand. For these results, we omit agnostic and EMP because at low density the performance is comparable.

\parab{Overtaking.}
\figref{fig:crash} shows, for \sysname and Single, a stacked bar that counts the number of outcomes of each type. Without cooperative perception, safe passage occurs only in 20\% of the configurations. Crashes occur in about 5\% of the configurations. These occur because the ego vehicle moves into the oncoming lane, but neither vehicle has enough time to stop. About 20\% of the configurations result in near misses, and the remaining settings result in deadlocks. By contrast, \sysname \textit{ensures safe passage  in all configurations}, but incurs near misses in about 5\% of the configurations.

\sysname's cooperative perception enables (\figref{fig:crash_detail}) much higher reaction times (average 13.3 seconds at 20~km/h), than without (average 0.31~s at 20~km/h, and zero at higher speeds), explaining its better performance. Finally, we have also investigated why \sysname incurs some near misses (with a minimum closest distance of 1.74 meters) in this scenario. At  low collider speeds, the ego is aware of the oncoming vehicle for 13 seconds and plans to stop the car as close and safe as possible to start lane change right after the collider passes; this represents a benign near-miss.
% thinks it is safe to move into the oncoming lane and starts to move a little by which time the collider arrives within the truck's LiDAR range. The ego stops, and the collider drives by, but at a very small separation.

% d dshows a bit more detail on the reaction time and the closest distance of these situations. Without \sysname, the reaction time for ego vehicle is 0, which implies the reason for an undesirable outcome is that it is too late for the ego vehicle to be aware of the collider. The occlusion severely limited the visibility which triggers a risky lane change decision. However, \sysname enables much earlier awareness through sensor sharing, increased the reaction time from 0 to upto 13.3 seconds when collider is at 20 km/h. The higher the collider speed, the lower reaction time gain. But the lowest gain is XXX sec, more than enough~\cite{twosecond} for the ego to make the right decision. \hangq{double check 30km/h data} The few near miss cases happen due to very high speed collider and limited lidar sensing range such that even with \sysname, the collider is not in the sensing range of the truck when the ego vehicle starts to make the lane change. More vantage points can solve this with \sysname.

\parab{Unprotected left-turn.}
This scenario is more benign than overtaking, because the ego is obstructed to a lesser extent. Without \sysname, \figref{fig:crash} shows that 16.7\% of the configurations resulted in a crash and 16.7\% in a near miss. For this scenario, \sysname ensures safe passage in all configurations. Because this scenario is benign, reaction times are in general higher both with and without \sysname. Without \sysname, crashes occur when the collider arrives in the shadow of the truck as the ego vehicle starts to turn left. When the collider's speed is high, it cannot brake fast enough to avoid the collision or near miss. With \sysname, reaction times and closest distances are generally quite high.

% t tThis can be demonstrated by the the longer average reaction time in the second column of \figref{fig:crash_detail}. The crash and near miss situation happens when the collider arrives exactly in the shadows of the truck when the ego vehicle starts the left turn. When they see each other, either the collider or the ego vehicle cannot brake in time to avoid the collision or near miss. \sysname is able to complete eradicate the risk in this scenario, giving full visibility to the truck's shadow for ego vehicle to track the collider's movement.

\parab{Red-light violation.}
%The red light violator scenario is another very dangerous scenario (see the middle column of \figref{fig:Scen10} for occlusion and visible area). Without \sysname, the ego vehicle will not be aware of the red light violator until it is very close to the last truck on the left lane. 
For this scenario, Single always incurs an undesirable outcome (75\% crashes, 25\% near miss). The occlusion angle is so wide that there is no time for the ego vehicle to react whatsoever (0 sec in the third column of \figref{fig:crash_detail}). By contrast, \sysname only incurs a few near-misses at low speeds: the controller is aware of the collider, but decides to leave little room to pass by.

\subsection{Pipeline Micro-benchmark}
\label{sec:microbenchmark}

\sysname carefully optimizes the perception and planning pipelines to achieve around 80 ms end-to-end processing latency. \figref{fig:pipelineeval} shows the average processing time and its variance for each module over 6000 frames, 2000 frames for each scenario. Object extraction and path planning are the most compute intensive. We use the Minkowski Engine~\cite{minkowski} for fast sparse quantization and Numba~\cite{numba} to speed up python loop execution such as the extended $A\star$ search and the isolated island detection. The execution time of each module may or may not be dependent on various factors. Quantization only depends on the number of lidar inputs. Creating the occupancy grid and extracting surrounding objects is depends on environment complexity, so their variances are high, but the maximum variance is below 20\%. Path planning takes on average 24.17 ms, but its results can be reused across frames with a fast waypoint check ($\sim$10 ms) across frames. The total latency is well below 100 ms, so \sysname can process LiDAR at full frame rate (10~fps). To sense and react to the dynamics in the environment, sensor suite (LiDar, Stereo Cameras, \etc) on autonomous vehicles are clocked at a minimum of 10~fps. Therefore it is critical to optimize the processing pipeline to operate over 10~fps.

% \begin{figure}
% \centering
% \centering
% \includegraphics[width=0.9\columnwidth]{fig/pipeline/pipelineeval.eps}
% \vspace{-5mm}
% \caption{Pipeline Micro Benchmark}
% \label{fig:pipelineeval}
% \end{figure}

\subsection{Experiments with V2V radios}
\label{sec:dsrc_exps}

\parab{Methodology.}
In this section, we replay the transmission schedules from one configuration of each scenario over a test bed with three DSRC radios; each transmission is carried over a random transceiver pair. We programmed the DSRC radios to use LTE-Direct TDMA Mode 4~\cite{LTE-Direct}, a listen-before-transmission mode to follow the schedule. To coordinate the radios, we designed simple handshake messages and timeout mechanisms to maintain synchronization among all radios. The precise synchronization mechanism that incurs minimum overhead is an open topic beyond the scope of this paper.

For each scenario, we record the point cloud data to transmit and the computed transmission schedule.  In the simulator, the schedule is based on a theoretical model of the channel with a fixed data rate. The goal of this  evaluation is to validate whether the DSRC radios can finish the scheduled transmission in time and evaluate the significance of packet loss and its impact on the transmissions.

\figref{fig:dsrc_latency} shows the end-to-end latency for each 100 ms decision interval. We played back traces with different traffic densities to see if DSRC radios can always fulfill the schedule. At low density, the number of objects to transmit is small\footnote{To evaluate DSRC under different bandwidth saturation level, we reduced the number of passive traffic participants in this setting.}. The bandwidth saturates at 20 vehicles and objects get prioritized: those over 100 ms is compensated next interval.  The results show that DSRC can always finish the schedule in time. Prioritization in the schedule is the reason \sysname can maintain high visibility on collider and safety-critical objects.
\figref{fig:scen6_dsrc}
% \figref{fig:Scen8_dsrc}, 
% and \figref{fig:Scen10_dsrc} 
takes the overtaking scenario as an example to show the transmissions of each object in detail without background traffic. The scheduled transmissions are in the upper subplot and the actual DSRC transmissions on the bottom. Each shared object is represented by a line with a unique color compared to other objects in the same decision interval. Each colored line starts from the time of the beginning of the transmission, ends when the transmission completes. The x-axis represents the time in ms, the y-axis represents the time within each decision interval (100~ms). If an object is lost due to channel variability and dropped packets, the solid line becomes a dashed line.
In the overtaking scenario, the collider's point cloud as observed by the truck is in yellow, blue represents the truck's point cloud as observed by the collider. As the collider moves closer to the truck, both observations are larger and hence the length of yellow and blue transmission is longer. Red is the ego's point cloud as observed by the truck, and purple is the truck's point cloud as observed by the ego. 
% In this regime, notice that the red lines are larger since the truck is always close to the ego vehicle, but the collider point cloud is much smaller (smaller yellow lines). 
Blue and purple transmissions are also scheduled: the ego and collider broadcasting the truck's (occluder's) point cloud. In theory, they need not be transmitted because the truck is visible and not relevant; \sysname does transmit objects with less relevance when possible. 

\begin{figure*}
\centering
\begin{minipage}{0.66\columnwidth}
\centering
\includegraphics[width=0.9\columnwidth]{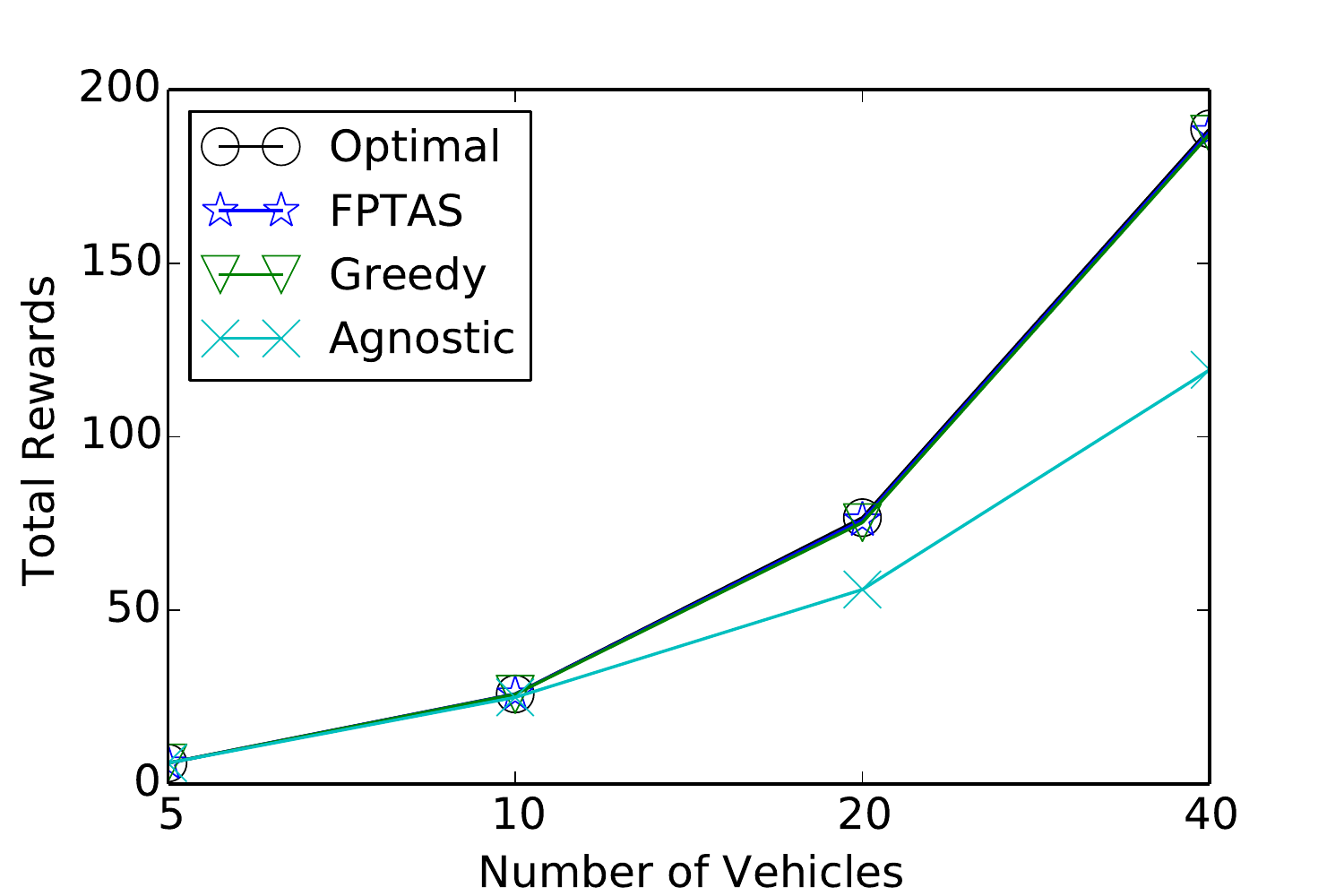}
% \vspace{-4mm}
\caption{Optimality: Total Rewards}
\vspace{-3mm}
\label{fig:scalability}
\end{minipage}
\hfill
\begin{minipage}{0.66\columnwidth}
\centering
\includegraphics[width=0.9\columnwidth]{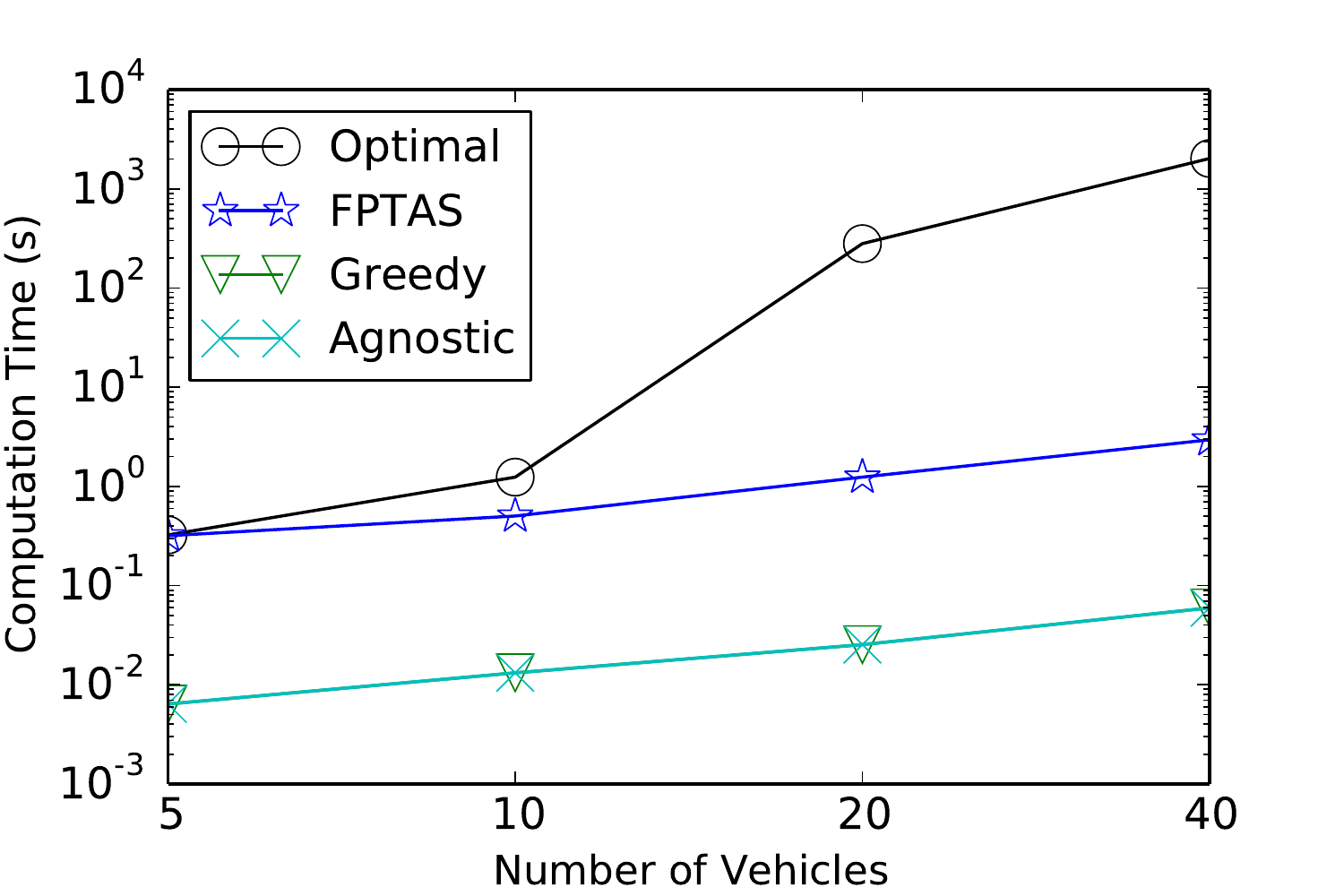}
% \vspace{-4mm}
\caption{Computation Time}
\vspace{-3mm}
\label{fig:exectime}
\end{minipage}
\hfill
\begin{minipage}{0.66\columnwidth}
\centering
\includegraphics[width=0.9\columnwidth]{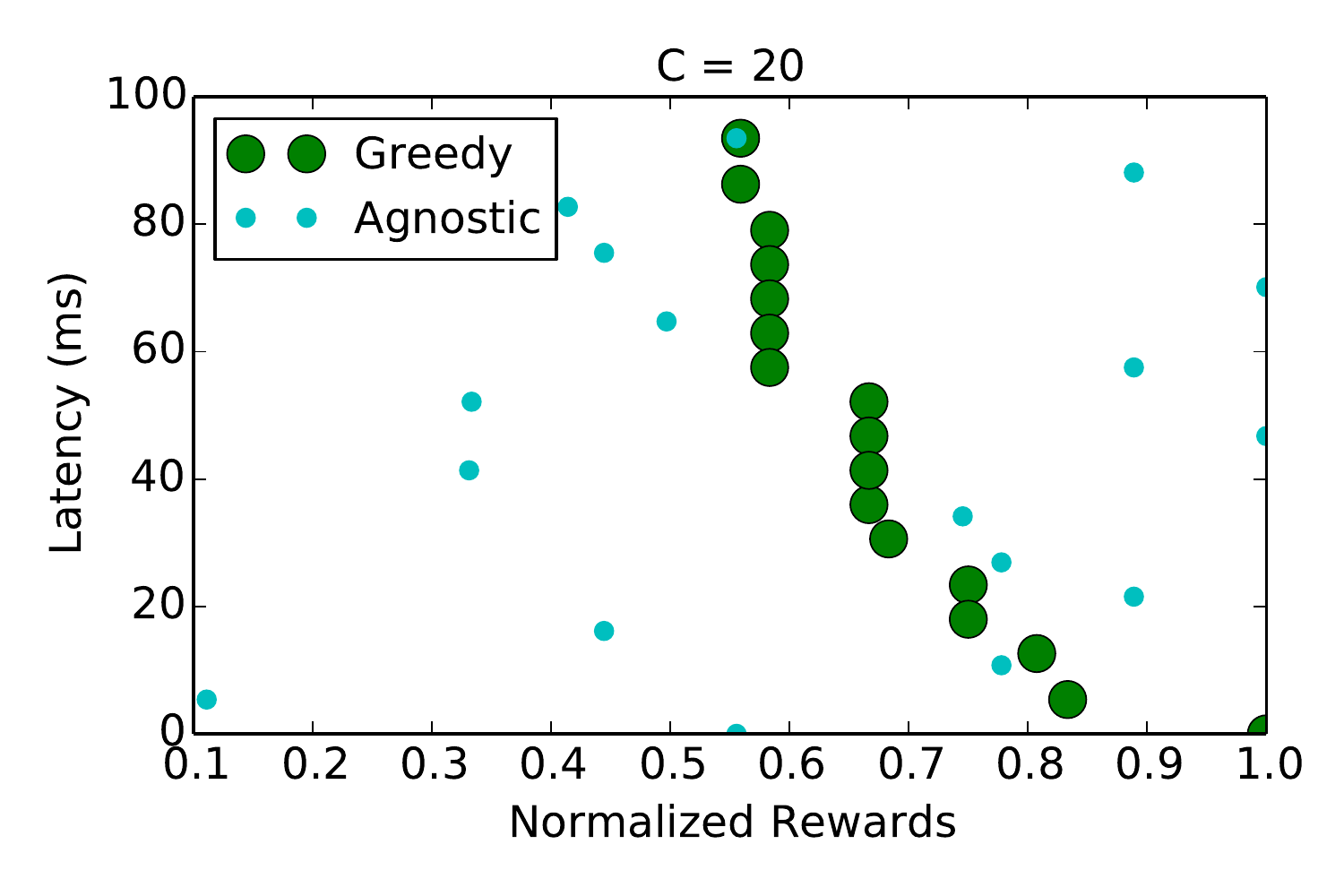}
% \vspace{-4mm}
\caption{Scheduled Delay}
\vspace{-3mm}
\label{fig:latencies}
\end{minipage}
\end{figure*}

\subsection{Scheduling Algorithms}
\label{sec:scheduler_eval}

We evaluate the optimality of different scheduling algorithm (Optimal, FPTAS, Greedy, Agnostic) in terms of total rewards and reward ratio, algorithm complexity and scalability with respect to the number of vehicles, and transmission delay for objects with different rewards. We conduct this set of by setting the vehicle on autopilot mode, entering and exiting an intersection from all directions. 

\parab{Optimality, Complexity, and Scalability.}
We first evaluate the optimality. \figref{fig:scalability} plots the total scheduled rewards when the number of vehicles varies from 5 to 40. Specifically, Greedy has less than 2\% difference from Optimal while Agnostic has upto 40\% difference from Greedy.  \figref{fig:exectime} further shows the computation time of the schedule with different number of vehicles. Although the total rewards are close, greedy is two orders of magnitude faster than FPTAS whereas the running time of optimal (dynamic programming) can quickly become prohibitive. Vehicular environments can be highly dynamic which requires the schedule to be computed frequently; only the proposed greedy algorithm can finish within the 100ms decision interval with 40 vehicles. 
% Finally, we study how much of the scheduled rewards can actually be received (reward ratio). The lower plot of \figref{fig:scalability} shows that 
% When bandwidth is sufficient (with 5 or 10 vehicles), all four algorithms can fulfill 100\% of the scheduled objects. 
% Also, it is interesting to see the reward starts to fall below 100\% when 20 vehicles are present. When the bandwidth is scarce, the optimality of \sysname is evident: \sysname achieves up to 25\% more reward while transmitting on average nearly 20\% more objects. 

% \begin{figure}
% \centering
% \includegraphics[width=\columnwidth]{fig/new/RR_time.eps}
% \vspace{-5mm}
% \caption{\textit{Reward Ratio} over Time (20 Vehicles).\hangq{we can maybe keep this figure and remove partitions.}}
% \label{fig:rr_time}
% \centering
% \vspace{-6mm}
% \end{figure}

% \figref{fig:rr_time} shows a detailed plot of the  \textit{reward ratio} metric over time. As vehicles join the intersection, and leave the intersection, the reward ratio changes for every \textit{sharing session} duration of 100 ms. 

\parab{Scheduled Delay.}
The scheduled delay is measured by calculating the duration from the start of each decision interval (every 100~ms) to the time that a particular object is received. \figref{fig:latencies} shows the scheduled delay of each object in a 20 cars scenario. It gives more details behind the scene, explains the reward ratio difference by showing the transmission priority.  \sysname's optimization always put the object with the highest normalized rewards top of the schedule which results in lower scheduled delay, whereas the latency of objects scheduled in agnostic is random. 

\section{Related Work} \label{ReWork}
\noindent
{\bf Connected Vehicles and Infrastructure:}
Connected vehicles promise great opportunities to improve the safety and reliability of self-driving cars. Vehicle-to-vehicle (V2V) and vehicle-to-infrastructure (V2X) communications both play an important role to share surrounding information among vehicles. 
Communication technologies, \eg, DSRC \cite{DSRC} and LTE-Direct \cite{LTE-Direct, Qualcomm:LTE-Direct}, provide capabilities to exchange information among cars by different types of transmission, \ie, multicasting, broadcasting, and unicasting. 
Automakers are deploying V2V/V2X communications in their upcoming models \cite{Mercedes, Toyota}). The academic community has started to build city-scale advanced wireless research platforms (COSMOS \cite{cosmos}), as well as large connected vehicle testbed in the U.S. (MCity \cite{USTest}) and Europe (DRIVE C2X \cite{EuropeTest}), which gives an opportunity to explore the application feasibility of connected vehicles via V2V communications in practice. 

\noindent
{\bf Sensor/Visual Information:}
Collecting visual information from sensors (\eg LiDAR, stereo cameras, \etc) is a major part of autonomous driving systems. 
These systems rely on such information to make a proper on-road decisions for detection \cite{Chen:CVPR}, tracking \cite{Ma:ICCV}, and motion forecasting \cite{Ma:CVPR}. In addition, there is a large body of work that explores vehicle context and behavior sensing \cite{MotionBased, InVehicle, qiu2018towards} to enhance vehicular situational awareness. 
Thanks to advanced computer vision and mobile sensing technologies, all the sensing capability can already be leveraged efficiently in a single car setting \cite{Luo:CVPR}. This work, and several related works discussed below, take the next step of designing how to share this information among nearby vehicles.

\noindent
{\bf Vehicle Sensor Sharing:}
Past research has attempted to realize some form of context sharing among vehicles. Rybicki et al. \cite{PeersOnWheels} discuss challenges of VANET-based approaches, and propose to leverage infrastructure to build distributed, cooperative Traffic Information Systems. Other work
has explored robust inter-vehicle communication~\cite{Cartalk,Kumar:SIGCOMM,corecast}, an automotive ontology for intelligent vehicles~\cite{automotiveOntology}, principles of community sensing that offer mechanisms for sharing data from privately held sensors~\cite{CommunitySensing}, and sensing architectures~\cite{semanticServices} for semantic services. Motivated by the ``see-through" system  \cite{Gomes:VNC}, several prior works on streaming video for enhanced visibility have been proposed \cite{Boban:VANET, Rameau:TVCG, Ferrerira:ISMAR, Judvaitis:ACCS, Lindemann:AutomotiveUI}. 
The above line of work has focused on scenarios with two cars, where a leader car delivers its whole video to a follower vehicle, which in many scenarios is not sufficient. 

Most recent work~\cite{emp, Qiu:Mobisys, chen2019fcooper} demonstrates the feasibility of sharing point clouds, but with limited scale and infrastructure support. In this work, we focus on enabling clusters of vehicles to share sensor information at scale in the absence of edge servers.
% we focus on scenarios with more than two cars, where multiple cars share a portion of their visual information with other cars via V2V communications. 
We design an end-to-end cooperative perception framework to compute when, which views, and to which cars to deliver views over time. 
While enabling on-demand cooperative perception, when infrastructure is available, our approach can naturally extend and work as-is. Because of the design of spatial reasoning and metadata exchange, as shown in \secref{sec:eval}, \sysname can scale better than existing baselines.
% In summary, the contributions of this work are as follows: we (i) partition 360 views into partial views and rank them based on their importance for other vehicles, (ii) create near optimal schedules to transmit partial views in decreasing order of importance, and (iii) consider compact representations of partial views to further pack more information into the limited bandwidth during the short timescales that we have.

%%% Local Variables:
%%% mode: latex
%%% TeX-master: "main"
%%% End: 
	
\section{Conclusion} \label{Conclusion}
In this paper, we have designed and implemented \sysname, a system that scales cooperative perception to dense traffic settings without infrastructure dependency. \sysname allows vehicles to share point clouds of dynamic objects with each other, but because these can congest the wireless channel, it carefully determines which objects to transmit based on visibility and relevance. These properties are input to a distributed scheduling algorithm that determines a transmission schedule at every decision interval. For several challenging traffic scenarios, \sysname significantly outperforms baseline approaches that do not employ cooperative perception, or do not prioritize objects transmissions, or rely on edge relay services. Its perception and planning pipelines have been optimized to process LiDAR data at frame rate in under 100~ms. Future work can improve perception and planning modules, experiment over LTE-V radios, experiment with using other representations for sharing, and design corresponding representation fusion for end-to-end control.

%This way, it can deliver up to 200\% more useful information to participating vehicles compared to a baseline algorithm which shares full views of vehicles in a round robin fashion. 
%%% Local Variables:
%%% mode: latex
%%% TeX-master: "main"
%%% End: 

\newpage
\balance
\bibliographystyle{abbrvnat}
\bibliography{references.bib}
\end{document}